\RequirePackage{rotating}
\documentclass[a4paper,fleqn,usenatbib]{mnras}

\usepackage[T1]{fontenc}
\usepackage{ae,aecompl}

\usepackage{graphicx}	
\usepackage{rotating}  
\usepackage{amsmath}	
\usepackage{amssymb}	
\usepackage{pifont}
\usepackage{caption} 

\newcommand{\cmark}{\ding{51}}%
\newcommand{\xmark}{\ding{55}}%
\newcommand{\com}[1]{\textnormal{#1}}
\newcommand{\comm}[1]{\textnormal{#1}}
\newcommand{\comme}[1]{\textnormal{#1}}

\title[Lensing distance ratio in a Friedmann universe]{Generalised model-independent characterisation of strong gravitational lenses V: reconstructing the lensing distance ratio by supernovae for a general Friedmann universe}

\author[J. Wagner and S. Meyer]{
Jenny Wagner$^{1}$\thanks{E-mail: j.wagner@uni-heidelberg.de}
and Sven Meyer,$^{2}$\thanks{E-mail: sven.meyer@uni-heidelberg.de}
\\
$^{1}$Universit\"at Heidelberg, Zentrum f\"ur Astronomie, Astronomisches Rechen-Institut, M\"onchhofstr. 12--14, 69120 Heidelberg, Germany\\
$^{2}$Universit\"at Heidelberg, Zentrum f\"ur Astronomie, Institut f\"ur Theoretische Astrophysik, Philosophenweg 12, 69120 Heidelberg, Germany\\
}

\date{Accepted XXX. Received YYY; in original form ZZZ}

\pubyear{2019}

\begin{document}
\label{firstpage}
\pagerange{\pageref{firstpage}--\pageref{lastpage}}
\maketitle

\begin{abstract}
\comm{We determine the cosmic expansion rate from supernovae of type Ia to set up a data-based distance measure that does not make assumptions about the constituents of the universe, i.e. about a specific parametrisation of a Friedmann cosmological model.} The scale, determined by the Hubble constant $H_0$, is the only free \comm{cosmological} parameter left \comm{in the gravitational lensing formalism}. 
We investigate to which accuracy and precision the lensing distance ratio $D$ is determined from the Pantheon sample. Inserting $D$ and its uncertainty into the lensing equations for given $H_0$, esp.\@ the time-delay equation between a pair of multiple images, allows to determine lens properties, esp.\@ differences in the lensing potential ($\Delta \phi$), without specifying a cosmological model. 
We expand the luminosity distances into an analytic orthonormal basis, determine the maximum-likelihood weights for the basis functions by a globally optimal $\chi^2$-parameter estimation, and derive confidence bounds by Monte-Carlo simulations. 
For typical strong lensing configurations between $z=0.5$ and $z=1.0$, $\Delta \phi$ can be determined with a relative imprecision of 1.7\%, assuming imprecisions of the time delay and the redshift of the lens on the order of 1\%. 
With only a small, tolerable loss in precision, the model-independent lens characterisation developed in this paper series can be generalised by dropping the specific Friedmann model to determine $D$ in favour of a data-based distance ratio. Moreover, for any astrophysical application, the approach presented here, provides distance measures for $z\le2.3$ that are valid in any homogeneous, isotropic universe with general relativity as theory of gravity. 
\end{abstract}




\section{Introduction}
\label{sec:introduction}

\subsection{Motivation from previous works}

There is hardly any astrophysical research question that does not involve the distance to an object of interest. Distance measurements in our galactic neighbourhood can be performed to high precision and accuracy, \cite{bib:Gaia}. \comm{Compared to that,} extra-galactic distance measurements \comm{are more} difficult, \cite{bib:Cuesta, bib:Muraveva,bib:Tegmark2002}. Thus, with the development of an observation-based cosmic distance ladder still ongoing, cosmic distances are usually inferred from a measured (spectroscopic or photometric) redshift in combination with a cosmological model that assigns the redshift to a cosmic distance. So far, the cosmological standard model, as most precisely measured by \cite{bib:Planck2018}, is inserted into these distance measures. 

As one example, in the gravitational lensing formalism, angular diameter distances between the observer and the lens $D_\mathrm{l}$, the observer and the source $D_\mathrm{s}$, and the distance between the lens and the source $D_\mathrm{ls}$ appear in the lensing equations to scale the (multiple) images, the source, and the deflection potential with respect to each other. The so-called lensing distance ratio
\begin{align}
D(z_\mathrm{l},z_\mathrm{s}) = \dfrac{D_\mathrm{l}D_\mathrm{s}}{D_\mathrm{ls}} \;,
\label{eq:D}
\end{align}
appears, for instance, in the time delay $\tau_{ij}$ between two multiple images $i$ and $j$ of the same background galaxy located at angular position $\boldsymbol{y}$ in the source plane, which is given by
\begin{equation}
\tau_{ij} = D \dfrac{(1+z_\mathrm{l})}{c} \left(\phi (\boldsymbol{y},\boldsymbol{x}_i) - \phi (\boldsymbol{y},\boldsymbol{x}_j )\right)  \equiv  D \dfrac{(1+z_\mathrm{l})}{c} \Delta \phi\;.
\label{eq:time_delay}
\end{equation}
$\boldsymbol{x}_i$ are the measured angular positions of the two images in the lens plane and $\phi(\boldsymbol{y},\boldsymbol{x})$ is the lensing potential; $c$ denotes the speed of light and $z_\mathrm{l}$ the redshift of the lens. The lensing potential is given by 
\begin{equation}
\phi(\boldsymbol{x},\boldsymbol{y}) = \dfrac12 \left( \boldsymbol{x} - \boldsymbol{y} \right)^2 - \psi(\boldsymbol{x}) \;,
\label{eq:lensing_potential}
\end{equation}
in which $\psi(\boldsymbol{x})$ denotes the projected, two-dimensional gravitational deflection potential of the lens in the lens plane. 
A detailed introduction of the gravitational lensing formalism can be found, e.g. in \cite{bib:Petters, bib:SEF}. 

In the previous four papers of this series, \cite{bib:Wagner1, bib:Wagner2, bib:Wagner3, bib:Wagner5}, we investigated gravitational lensing from a model-independent perspective. We derived equations that determine local lens properties for different configurations of multiple images solely from observed properties of these images without assuming a specific model for the lens and determined invariance transformations of these equations to derive the degeneracies in our approach. As the next step of generalisation of this ansatz, we now replace the distance measures \com{defined in Equations~\eqref{eq:D_L} and \eqref{eq:D_A}} based on a cosmological model by data-based distance measures that do not involve a particular parametrisation of a Friedmann-Robertson-Lemaître-Walker cosmological model. We call  the derived distances therefore Friedmann-parameter-free in the following. \com{This enables us to determine $\Delta \phi$ without the need to specify in which way the total energy density of our universe today is distributed among radiation, matter, curvature, a cosmological constant, or dark energy.} Since most lenses and their background sources are located at redshifts between $z=0$ to $z=2.0$, reconstructing their distances by a supernova sample, like the Pantheon sample, \cite{bib:Scolnic}, is possible. Hence, we can obtain data-based distances for most lensing configurations without the need to calibrate several probes of the cosmic expansion with respect to each other. 

The paper is organised as follows: in the remainder of this section, we discuss related work on the usage of supernovae to reconstruct the cosmic expansion history \comm{and highlight the differences between model fits and data-driven reconstructions of the cosmic expansion history}. Then, Section~\ref{sec:theoretical_derivations} gives a brief introduction into cosmological model building based on Friedmann universes and discusses the observational prerequisites and basics of the standardisation of supernovae that influence the reconstructions of the cosmic expansion and distance measures. \com{The Pantheon sample (contrary to its antecessor\comm{, the UnionSample 2.1, \cite{bib:Suzuki}}) is not calibrated by an overall scale. Therefore, we derive the equations to reconstruct the expansion function from a scale-free sample by normalising the expansion function. Subsequently, we insert this normalised expansion function into the definitions of distance measures.} In Section~\ref{sec:implementational_details}, \comm{details about the implementation of} the approach are given, before it is applied to the Pantheon sample and a Pantheon-like simulation in Section~\ref{sec:application}. 
Since $H_0$ is not determined from the scale-free supernovae data, it remains a free parameter that has to be set independently. We insert $H_0$ as derived from cosmic-microwave background measurements or as measured in our local neighbourhood into the distance measures. Together with the expansion function from the supernovae, we obtain data-based luminosity and angular diameter distances. We compare both choices for $H_0$ and show their deviations in the resulting distances. Independent of $H_0$, we evaluate the relative accuracy and  precision of our implementation to reconstruct the expansion function and the distances up to this overall scaling by $H_0$. In addition, we compare our Friedmann-parameter-free reconstructions of the expansion function, the luminosity distances, and the lensing distance ratio to the ones of $\Lambda$ cold dark matter ($\Lambda$CDM) models, as parametrised by \cite{bib:Planck}\footnote{being the most recent data release at the time of performing our experiments} and \cite{bib:Scolnic}. \comm{As a consistency check, we also perform a fit of the data from the Pantheon sample to a flat $\Lambda$CDM cosmology with our $\chi^2$-parameter-estimation function to show that our mathematical framework leads to a Friedmann $\Omega_{\mathrm{m}0}$-parameter value that is in agreement with the value for $\Omega_{\mathrm{m}0}$ as obtained by \cite{bib:Scolnic} with a different parameter-estimation approach on the same dataset and that is in agreement with the value obtained by \cite{bib:Planck}.} In Section~\ref{sec:synopsis}, we compare the precision of the lensing distance ratio to the precision of the other observables entering the time-delay equation of two multiple images of the same background source to estimate the loss of precision when dropping the parametrisation of the Friedmann model in favour of a data-based distance measure. 
Section~\ref{sec:conclusion} summarises the results, discusses the advantages and disadvantages of Friedmann-parameter-free distance measures and gives an outlook for further applications.
\com{This work focuses on the development of the Friedmann-parameter-free distance measures to determine $\Delta \phi$ for a fixed $H_0$. Determining $H_0$ for a given $\Delta \phi$ will be the subject of the next part of the paper series.}

\subsection{Methodology and related work of supernovae of type Ia as cosmological probes}

The idea to reconstruct the evolution of the cosmic density (perturbations) from luminosity distances of standardisable candles goes back to \cite{bib:Starobinski} and has been employed in many ways and variants since then. The basic ansatz solves
\begin{align}
D_\mathrm{L}(a) &= \dfrac{c}{a} \int \limits_{a}^{1} \dfrac{\mathrm{d}x}{x^2 H(x)} 
\label{eq:starobinski}
\end{align}
for $H(a)$ having measured the luminosity distance $D_{L}(a)$ from the observer to supernovae at the left-hand side and assuming that the universe is flat. 
\cite{bib:Tegmark2002} extended and detailed the original idea and suggested combinations of different probes of the cosmic density. \cite{bib:Huterer} provided first feasibility tests for different dark energy potentials by Monte-Carlo simulations.

Depending on their way, how to invert Equation~\eqref{eq:starobinski}, several approaches are distinguished. \cite{bib:Shafieloo1} contains a detailed list of approaches, therefore, we focus on more recent ones here: Direct methods reconstruct $H(a)$ from the smoothed or averaged version of the original data set, e.g. as done in \cite{bib:Shafieloo1, bib:Shafieloo2, bib:Wang2005}. Basis-function methods reconstruct $H(a)$ by expanding $D_\mathrm{L}(a)$, the distance modulus $\mu$, $H(a)$, or a parametrisation thereof, into a set of basis functions, e.g. as done in \cite{bib:Gomez, bib:Ishida, bib:Mignone}. Bayesian methods reconstruct $H(a)$ either employing Gaussian processes or other priors, e.g. as done in \cite{bib:Gomez, bib:Porqueres, bib:Seikel3}, and the most recently by \cite{bib:Lemos} (and references therein) \comm{and \cite{bib:Capozziello}}.

In addition, other ansatzes exist that use supernovae of type Ia to establish a model-independent hypothesis test to reject the hypothesis that the universe is not expanding, e.g. like \cite{bib:Seikel1, bib:Seikel2}. 
\comm{Another hypothesis test can be found in \cite{bib:Zhao}. It is based on the Kullback-Leibler divergence and uses supernovae of type Ia in combination with other cosmological probes to find tensions between the data sets and in the current cosmological standard model.}

\comm{Reconstructing $H(a)$ by fitting parametric models, e.g. a $\Lambda$CDM model, to the data, yields the most likely parameter values together with their confidence bounds. Due to the limited amount of parameters, parametric models can become unlikely for an increasing amount of data or with increasing measurement precision. For instance, the flat $\Lambda$CDM model can explain the cosmic evolution as measured by the Pantheon data set but may require extensions for future supernovae data of higher measurement precision or for increasing redshift. Extending a parametric model like the flat $\Lambda$CDM model is not unique and introducing further parameters is motivated by adding physical assumptions. As a disadvantage, the resulting model may not be the only feasible solution and may have intrinsic degeneracies among the parameters (as shown e.g.\@ in \cite{bib:Planck}). Parametric models thus yield a description for $H(a)$ that can be directly interpreted in physical terms and that tests the compatibility of our assumptions with observations. They are the most suitable ansatz to reconstruct $H(a)$ when only sparse observational data is available.}

\comm{Complementary to parametric model fitting, $H(a)$ can be reconstructed from standardisable observational data. This ansatz yields a flexible representation of $H(a)$ that is uniquely extended for an increasing amount of data. The disadvantage of data-driven reconstructions is often the data representation itself that, like in case of basis functions, lacks a direct physical interpretation. The width of the confidence bounds to which the cosmic evolution can be constrained by the data is a measure of the constraining power of these data. The wider the confidence bounds of the reconstruction, the less constraining the data are. Usually, parametric model fits have much smaller confidence bounds due to the additional assumptions they make. With an increasing amount of data becoming available, data-driven methods will overcome their disadvantage of being less precise than model-based methods and observations will replace model assumptions. Considering the supernovae of type Ia available in \cite{bib:Suzuki} and in the Pantheon sample of \cite{bib:Scolnic}, the amount of data has almost doubled within six years and the measurement precision and the redshift range have also been increased.} 

\comm{In our approach to determine distance measures from a set of supernovae of type Ia, we choose a data-driven reconstruction based on statistical sampling theory as an alternative to the most commonly used parametric models. As data representation, we employ the orthonomal set of basis functions as developed in \cite{bib:Mignone}, which is physically motivated. This set of basis functions has been further investigated in \cite{bib:Benitez12, bib:Benitez13}. From all distance measures mentioned above, it seems to be most suitable for the reconstruction of the lensing distance ratio $D$ in our approach to gravitational lensing, as will be further detailed in Section~\ref{sec:theoretical_derivations} and \ref{sec:implementational_details}, and tested in Section~\ref{sec:application}.}
 
\section{Theoretical derivations}
\label{sec:theoretical_derivations}

\subsection{Cosmological model prerequisites}
\label{sec:cosmology}
Assuming that our universe is spatially homogeneous and isotropic on large scales and general relativity is the theory of gravitation, the universe can be described by a Friedmann-Lemaître-Robertson-Walker metric (FLRW metric), which fulfils the first order Friedmann equation,  
\begin{align}
H(t)^2 \equiv \left( \dfrac{\dot{a}(t)}{a(t)} \right)^2 = \dfrac{8\pi G}{3} \rho - \dfrac{Kc^2}{a^2} + \dfrac{\Lambda c^2}{3} \;,
\end{align}
in which $G$ denotes the gravitational constant, \comm{$\rho$ is the matter-radiation energy density}, $K$ is the constant spatial curvature, and $\Lambda$ the cosmological constant. The dot denotes the derivative with respect to cosmic time $t$. Without loss of generality, we choose the scale factor today to be one, $a_0 \equiv a(t_0) =1$, in order to determine $a$ uniquely.

Being agnostic about the constituents and the state of the universe (i.e.\ not knowing anything about $\rho, K$, or $\Lambda$), we can generally define 
\begin{align}
H(a) \equiv H_0 E(a) \;,
\label{eq:H}
\end{align}
calling $H(a)$ the Hubble function of the universe with the expansion function $E(a)$ and today's Hubble constant $H_0=H(a_0)$. Thus, $E(a)$ is normalised, such that $E(a_0)=1$. \comm{This is the most general functional form of $H(a)$ to describe the evolution of a spatially homogeneous and isotropic metric fulfilling Einstein's field equations.} Splitting the constituents into radiation, matter, curvature and a $\Lambda$-term, we arrive at the usual parametrisation of the Hubble function
\begin{align}
H(a) = H_0 \sqrt{\Omega_{r0}a^{-4} + \Omega_{m0} a^{-3} + \Omega_K a^{-2} + \Omega_\Lambda} \;,
\label{eq:H_sm}
\end{align}
with $\Omega_i$ being today's density divided by the critical density.

With the help of Equation~\eqref{eq:H} and the distance duality relation of \cite{bib:Etherington}, we can calculate the luminosity and the angular diameter distances between two scale factors $a_1$ and $a_2$ with $a_2 < a_1$, as
\begin{align}
D_\mathrm{L}(a_1,a_2) &= \dfrac{c}{H_0} \dfrac{a_1}{a_2} f_K\left( \int \limits_{a_2}^{a_1} \dfrac{\mathrm{d}x}{x^2 E(x)} \right) \;, \label{eq:D_L} \\
D_\mathrm{A}(a_1,a_2) &= \dfrac{c}{H_0} \dfrac{a_2}{a_1} f_K\left( \int \limits_{a_2}^{a_1} \dfrac{\mathrm{d}x}{x^2 E(x)} \right) = \left( \dfrac{a_2}{a_1} \right)^2 D_\mathrm{L}(a_1,a_2) \label{eq:D_A} \;,
\end{align}
with 
\begin{align}
f_K(r) = \left\{ \begin{matrix} \dfrac{\sinh(\sqrt{|K|}r)}{\sqrt{|K|}} & \text{for} \; K < 0 \\ r & \text{for} \; K=0 \\ \dfrac{\sin(\sqrt{K}r)}{\sqrt{K}}  & \text{for} \; K > 0 \end{matrix} \right. \;.
\label{eq:f_K}
\end{align} 

\subsection{Observational prerequisites}
\label{sec:observational_properties}

In order to employ Equations~\eqref{eq:D_L} and \eqref{eq:D_A} to reconstruct $H(a)$, observations from standardisable candles, rulers, or sirens can be used. For supernovae of type Ia, determining $D_\mathrm{L}$ by fitting their light curves to standardised templates has become a routine (see e.g.\ \cite{bib:Amanullah, bib:Betoule, bib:Burns, bib:Scolnic, bib:Suzuki}). Light curve fitters, like SALT2 (\cite{bib:Guy}), determine \comm{the distance modulus $\mu$, which is a function of the observed light curves (usually measured as magnitudes in the $b$-band). The light curves are parametrised by the peak magnitude, the time-stretching of the supernova, its colour at maximum brightness, the absolute magnitude $M$ for a standardised supernova of type Ia, and parameters of the light curve fitter to correct for biases, e.g.\ in the distance modulus due to the host-galaxy mass of the individual supernovae.} The impact of these biases \comm{can be} calibrated by simulations that may depend on an underlying Friedmann model with a special parametrisation. Thorough analyses have been performed that investigate the dependence of the inferred quantities, e.g. $\mu$ or cosmological parameters, on the trained and calibrated light-curve model, see \cite{bib:Mosher} and \cite{bib:Hauret}. Enhanced simulations that employ several parametrised Friedmann models for the calibration and blind the light curve fitting with respect to a reference cosmology, have also been established, \cite{bib:Kessler}.
\comm{In the case of the Pantheon sample, the absolute magnitude $M$ of a standard supernova is left as a free parameter. Given $\mu$, we can determine the luminosity distance}
\begin{align}
D_\mathrm{L} = 10^{\tfrac{\mu}{5}+1} \;.
\label{eq:D_Lmu}
\end{align}

Setting Equation~\eqref{eq:D_Lmu} equal to \eqref{eq:D_L} and inserting Equation~\eqref{eq:H_sm} into the latter, cosmological parameter values can be retrieved. Usually, nuisance parameters like the absolute magnitude $M$ are fitted together with the cosmological parameters\footnote{Determing $M$ for different, redshift-dependent subsets, as performed in \cite{bib:Suzuki}, revealed that all $M$ are in good agreement with each other. Thus, a single scale $M$ for the entire data set is sufficient.}. $M$ and $H_0$ both define an overall scale for the distances and are thus not independent of each other. Since the light-curve standardisation is performed without fixing $M$, observations of supernovae only determine $E(a)$ without constraining the overall distance scale. This has to be set by one of the following options:
\begin{itemize}
\item scaling with \com{an independently determined} $H_0$, e.g. as derived from \cite{bib:Planck} or from \cite{bib:Riess_H0}, or
\item a measurement of $M$ for supernovae in our local neighbourhood, \cite{bib:Richardson}, or
\item employing Bayesian statistics to infer cosmological parameter values after having marginalised over the prior probability of all nuisance parameters, e.g.\ assuming a non-informative prior probability distribution for $M$, \comm{see e.g.\@ \cite{bib:Conley} and \cite{bib:Scolnic}} (which is not pursued in this work).
\end{itemize}
 
In our work, we use the most recent Pantheon sample from \cite{bib:Scolnic}. It provides $\mu + M$, i.e. the distance to the supernovae up to an overall scaling factor, from a compilation of $N_\mathrm{D} = 1048$ supernovae. \comm{These were observed between} $a_\mathrm{min}=0.307$ \comm{($z=2.26$)} and $a_\mathrm{max}=0.990$ \comm{($z=0.01$), which set the limits in $a$ (and $z$) in which our distance measure can be used}. Basing on the frequentist framework of statistics for the remainder of this work, we reconstruct the normalised expansion function, $E(a)$, from these \com{scale-free} supernovae and subsequently scale with $H_0$ from \cite{bib:Planck}. 
\comm{Scaling by $H_0$ instead of using a local value for $M$} seems to be the most consistent approach because observations of local supernovae properties might be subject to small-scale anisotropy biases that are not included in the FLRW metric from which we derive $E(a)$ and the distance measures. See e.g.\ \cite{bib:Planck} for a detailed discussion about the tension of currently available measurements of $H_0$ and \cite{bib:Marra}, \cite{bib:Bolejko}, and \cite{bib:Macpherson} for recent advances to reconcile the measurements from the cosmic microwave background and the local neighbourhood.

\subsection{Scale-free series expansion of $D_\mathrm{L}$}
\label{sec:taylor_series}

\com{In a first step, we expand the scale-free distances to the supernovae into a set of basis functions:}
We define the luminosity distance function $D_\mathrm{L}(a,\boldsymbol{c})$ as an expansion into orthonormal basis functions $\phi_\alpha(a)$
\begin{align}
D_\mathrm{L}(a,\boldsymbol{c}) = \sum \limits_{\alpha=0}^{N_\mathrm{B}-1} c_\alpha \phi_\alpha(a) = \boldsymbol{c} \circ \Phi
\label{eq:series_expansion}
\end{align}
for $a \in \left[ a_\mathrm{min}, 1 \right]$, in which the $c_\alpha$ denote the weights of the basis functions and each entry in $\boldsymbol{c} \in \mathbb{R}^{N_\mathrm{B}}$ is multiplied by the respective column in $\Phi \in \mathbb{R}^{N_\mathrm{D} \times N_\mathrm{B}}$ and these terms are summed up in the short-hand notation of the right hand side. This ansatz has also been pursued in \cite{bib:Benitez12, bib:Benitez13, bib:Mignone}. Since, apart from some minor restrictions detailed in Appendix~\ref{app:validity}, $D_\mathrm{L}(a,\boldsymbol{c})$ can be exactly represented in any basis \comm{(for $N_\mathrm{B} \rightarrow \infty$)}, we do not insert a specific one until Section~\ref{sec:onb}  \comm{and establish a quality measure to compare different bases with respect to their practical usefulness in Section~\ref{sec:quality_assessment}}.

Denoting the data as provided by \cite{bib:Scolnic} as $d_i \equiv d(a_i) = \mu_i + M$, we factor out the unknown overall scale in Equation~\eqref{eq:D_Lmu} as
\begin{align}
D_{\mathrm{L},i} = 10^{\tfrac{\mu_i}{5}+1} = 10^{\tfrac{d_i-M}{5}+1} \equiv 10^{-\tfrac{M}{5}} \tilde{D}_{\mathrm{L},i} \;, \quad \forall i=1,...,N_\mathrm{D} \;.
\label{eq:D_Lred}
\end{align}

In order to link Equations~\eqref{eq:series_expansion} and \eqref{eq:D_Lred}, we define
\begin{align}
D_\mathrm{L}(a,\boldsymbol{c}) = 10^{-\tfrac{M}{5}} \tilde{D}_\mathrm{L}(a,\tilde{\boldsymbol{c}})  = 10^{-\tfrac{M}{5}} \tilde{\boldsymbol{c}} \circ \Phi\;.
\label{eq:series_expansion_red}
\end{align}

Given the covariance matrix between the $d_i$ (including systematic correlations), $\Sigma_\mu \in \mathbb{R}^{N_\mathrm{D} \times N_\mathrm{D}}$, we obtain the entries of the scale-free covariance matrix for $D_\mathrm{L}$, $\tilde{\Sigma}$, by calculating
\begin{align}
\Sigma_{ij} &= D_{\mathrm{L},i} D_{\mathrm{L},j} \, k_\Sigma = 10^{-\tfrac{2M}{5}} \tilde{D}_{\mathrm{L},i} \tilde{D}_{\mathrm{L},j} \, k_\Sigma \label{eq:sigma_ij} \\
&\equiv 10^{-\tfrac{2M}{5}} \tilde{\Sigma}_{ij} \;, \quad \forall i,j=1,...,N_\mathrm{D} \label{eq:sigma}
\end{align}
with
\begin{align}
k_\Sigma = 10^{\tfrac{\Sigma_{\mu,ij}}{5\sqrt{\Sigma_{\mu,ii}}} + \tfrac{\Sigma_{\mu,ij}}{5\sqrt{\Sigma_{\mu,jj}}}} - 10^{\tfrac{\Sigma_{\mu,ij}}{5\sqrt{\Sigma_{\mu,ii}}}} - 10^{\tfrac{\Sigma_{\mu,ij}}{5\sqrt{\Sigma_{\mu,jj}}}} +1 \;.
\end{align}
A derivation of $\Sigma_{ij}$ can be found in Appendix~\ref{app:sigma_derivation}. 

To obtain the $\boldsymbol{c}$ in Equation~\eqref{eq:series_expansion} from Equation~\eqref{eq:D_Lred}, we set up a generalised linear-least-squares parameter estimation as
\begin{align}
\mathrm{arg} \min \limits_{{\boldsymbol{c}}} \chi^2 \;,
\label{eq:chi2}
\end{align} 
with
\begin{align}
\chi^2 = \left( \boldsymbol{D}_\mathrm{L} - \boldsymbol{D}_{\mathrm{L}}(a,\boldsymbol{c})\right)^\top \Sigma^{-1} \left( \boldsymbol{D}_\mathrm{L} - \boldsymbol{D}_{\mathrm{L}}(a,\boldsymbol{c})\right) \;,
\end{align}
in which $\boldsymbol{D}_\mathrm{L} = \left(D_{\mathrm{L},1}, ..., D_{\mathrm{L},N_\mathrm{D}} \right)^\top$ denotes the column vector of the luminosity distance measurements and $\boldsymbol{D}_{\mathrm{L}}(a,\boldsymbol{c})$ is the column vector containing the luminosity distances at the same $a$ as determined by Equation~\eqref{eq:series_expansion}\footnote{This ansatz assumes that the errors in $a$ are negligible, which is realised by incorporating the uncertainties and biases in the redshift measurements in the covariance matrix $\Sigma_\mu$, see \cite{bib:Betoule}.}.
\\
Inserting Equations~\eqref{eq:D_Lred}, \eqref{eq:series_expansion_red}, and \eqref{eq:sigma} into Equation~\eqref{eq:chi2}, we obtain
\begin{align}
\chi^2 & = \left(\tilde{\boldsymbol{D}}_{\mathrm{L}} - \tilde{\boldsymbol{D}}_{\mathrm{L}}(a,\tilde{\boldsymbol{c}})\right)^\top \tilde{\Sigma}^{-1} \left( \tilde{\boldsymbol{D}}_{\mathrm{L}} - \tilde{\boldsymbol{D}}_{\mathrm{L}}(a,\tilde{\boldsymbol{c}})\right) \label{eq:chi2_red} \;.
\end{align}
Hence, the scaled and scale-free optimisation problems, and thus, their solutions, are of the same form and yield the unbiased, consistent, efficient, and asymptotically normal generalised least-squares estimator\footnote{under the assumption that $\mathbb{E}\left[\boldsymbol{D}_{\mathrm{L}} - \boldsymbol{D}_\mathrm{L}(a,\boldsymbol{c}) \, |\, \boldsymbol{D}_\mathrm{L}\right] = 0$}
\begin{align}
\hat{\boldsymbol{c}} &= \left( \Phi^\top \Sigma^{-1} \Phi \right)^{-1} \left(\Phi^\top \Sigma^{-1} \right) \boldsymbol{D}_\mathrm{L} = 10^{-\tfrac{M}{5}} \hat{\tilde{\boldsymbol{c}}} \;.
\label{eq:MLE}
\end{align} 
If the deviations of $\boldsymbol{D}_\mathrm{L}$ to $\boldsymbol{D}_\mathrm{L}(a,\boldsymbol{c})$ are normally distributed, it is also the maximum-likelihood estimator. For the sake of convenience, unless specified otherwise, we will drop the tilde and refer to the scale-free solution as $\hat{\boldsymbol{c}}$.

Compared to other methods that expand $E(a)$ in a set of basis functions, the ansatz pursued in Equation~\eqref{eq:series_expansion} has the advantage that it is easy to show that Equation~\eqref{eq:chi2} has a single global optimum which can be efficiently determined by Equation~\eqref{eq:MLE}.

\subsection{Reconstruction of the Hubble function}
\label{sec:e}

\com{In the next step, we use the basis function expansion of the scale-free supernovae distances to determine $E(a)$. By construction, $E(a)$ is determined up to an overall scale. Therefore, we employ the normalisation condition $E(a=1)=1$, such that the overall scale of $H(a)$, and the consequently the scale of the distance measures in Equations~\eqref{eq:D_L} and \eqref{eq:D_A}, is given by $H_0$. Hence, $H_0$ remains the only free parameter left in Equations~\eqref{eq:D_L} and \eqref{eq:D_A}.}

Luminosity distances of supernovae are measured with respect to $a=1$ today. Hence, when inserting $D_\mathrm{L}(a,\hat{\boldsymbol{c}})$ on the left-hand side of Equation~\eqref{eq:D_L}, we also have to insert $a_1 = 1$ and $a_2 = a$ on the right-hand side. If not explicitly specified, we abbreviate $D_\mathrm{L}(a)\equiv D_\mathrm{L}(1,a)$ for the luminosity distance defined by Equation~\eqref{eq:D_L} and analogously for the angular diameter distance measure.

Following \cite{bib:Starobinski}, we can solve Equation~\eqref{eq:D_L} for $E(a)$ by first isolating the integral of the right-hand side
\begin{align}
\int \limits_{a}^{1} \dfrac{\mathrm{d}x}{x^2 E(x)} = f^{-1}_K\left( \dfrac{H_0}{c} a D_\mathrm{L}(a) \right) \;,
\label{eq:integral_equation1}
\end{align}
subsequently deriving both sides by $a$
\begin{align}
-\dfrac{1}{a^2 E(a)} = \dfrac{H_0}{c} \dfrac{\partial f^{-1}_K\left( \tfrac{H_0}{c} a D_\mathrm{L}(a) \right)}{\partial \left( \tfrac{H_0}{c} a D_\mathrm{L}(a) \right)} \left( D_\mathrm{L}(a) + a \dfrac{\mathrm{d}D_\mathrm{L}(a)}{\mathrm{d}a} \right)
\label{eq:integral_equation2}
\end{align}
and then obtain $E(a)$ as
\begin{align}
E(a) = - \left[ a^2 \dfrac{H_0}{c} \dfrac{\partial f^{-1}_K\left( \tfrac{H_0}{c} a D_\mathrm{L}(a) \right)}{\partial \left( \tfrac{H_0}{c} a D_\mathrm{L}(a) \right)} \left( D_\mathrm{L}(a) + a \dfrac{\mathrm{d}D_\mathrm{L}(a)}{\mathrm{d}a} \right) \right]^{-1} \;.
\label{eq:E}
\end{align}
We restrict the discussion to flat universes with $K=0$, so that $f_K(r) = r$, which is in agreement with the measurements of \cite{bib:Planck}. 
The cases for non-vanishing curvature are analogous and treated in Appendix~\ref{sec:k} for the sake of completeness.

Inserting $D_\mathrm{L}(a,\hat{\boldsymbol{c}})$ for $D_\mathrm{L}(a)$ into Equation~\eqref{eq:E}, we obtain the expansion function up to an overall scale, $\tilde{E}(a)$. If the supernova sample were calibrated such that $c/H_0 = 10^{-M/5}$, the reconstructed expansion function would be normalised, $\tilde{E}(a) = E(a)$. \com{Yet, the Pantheon sample is scale-free, i.e. not calibrated by the Hubble scale $H_0$, nor by an absolute magnitude $M$, so that we have to insert the data-based scale-free series expansion $\tilde{D}_\mathrm{L}(a,\hat{\boldsymbol{c}})$ into Equation~\eqref{eq:E}. In this way, we obtain a normalised expansion function by}
\begin{align}
E(a) = \dfrac{\tilde{E}(a)}{\tilde{E}(a_\mathrm{max})} \;.
\end{align}

Inserting $\tilde{D}_\mathrm{L}(a,\hat{\boldsymbol{c}})$ and its derivative into Equation~\eqref{eq:E}, dropping all scale factors, and normalising the result, $E(a)$ is given by
\begin{align}
E(a) = - \left[ a^2 \tilde{E}(a_\mathrm{max}) \; \hat{\boldsymbol{c}} \circ \left( \Phi + a \dfrac{\mathrm{d}\Phi}{\mathrm{d}a} \right) \right]^{-1} \;.
\label{eq:E_full}
\end{align}
Consequently, when introducing the overall scale to determine $H(a)$, we use
\begin{align}
H(a) = H(a_\mathrm{max}) \; E(a) \;.
\label{eq:H_scaled}
\end{align}
Since, for the Pantheon sample, $a_\mathrm{max} =0.99 \approx 1$ and $H(a_\mathrm{max})=68.06$ km/s/Mpc for the Planck cosmological parameters (\cite{bib:Planck}) is contained in the confidence interval of $H_0 = 67.74 \pm 0.46$ km/s/Mpc, we may also approximate $H(a_\mathrm{max})$ by $H_0$ in Equation~\eqref{eq:H_scaled}. 

Since the basis functions can be determined at all $a \in \left[a_\mathrm{min}, 1\right]$ (either analytically or numerically), Equations~\eqref{eq:E_full} and \eqref{eq:H_scaled} yield a data-based expansion and Hubble function at any $a \in \left[a_\mathrm{min}, 1\right]$.

\subsection{Reconstruction of distances}
\label{sec:dl}

Inserting Equation~\eqref{eq:H_scaled} into Equations~\eqref{eq:D_L} or \eqref{eq:D_A}, we can now determine distances between arbitrary scale factors without specifying a parametrisation like in Equation~\eqref{eq:H_sm}.

If an unbiased, global measurement of an absolute distance for a standardised supernova $M$ were available, we could simply insert $M$ into Equation~\eqref{eq:series_expansion_red} and solve Equation~\eqref{eq:chi2} for $\hat{\boldsymbol{c}}$ (i.e. the scaled quantity as defined on the left-hand side of Equation~\eqref{eq:MLE}) to determine data-based distance measures. As we will briefly show in Section~\ref{sec:accuracy}, this would lead to a higher degree of accuracy and tighter confidence bounds. Until such a measurement might become feasible, Sections~\ref{sec:precision} and \ref{sec:accuracy} show that employing Equations~\eqref{eq:D_L} or \eqref{eq:D_A} with \eqref{eq:H_scaled} currently is the most robust way to determine data-based distances. \com{Recent progress on the determination of $M$ can be found in \cite{bib:Richardson}.}

\subsection{Reconstruction of the lensing distance ratio}
\label{sec:d}

With the establishment of data-based distance measures as detailed in Section~\ref{sec:dl}, we can determine the lensing distance ratio for all lenses and sources located within the redshift\footnote{While we use $a$ as variable for the reconstructions of the Hubble function and the distance measures, we convert $a$ to the redshift $z$ for the reconstruction of $D$ to be consistent with standard lensing notation.} range of the reconstruction, as defined in Equation~\eqref{eq:D}. In Sections~\ref{sec:precision} and \ref{sec:accuracy}, we will specify a lens redshift $z_\mathrm{l}$ and calculate the lensing distance ratio for all sources at $z_\mathrm{s} > z_\mathrm{l}$.

\section{Implementational details}
\label{sec:implementational_details}

\subsection{Analytic bases}
\label{sec:onb}

\subsubsection{Einstein-de-Sitter basis}
\label{sec:EdS}

As a set of orthonormal basis functions, we use the Einstein-de-Sitter basis as introduced in \cite{bib:Mignone}. It consists of functions
\begin{align}
u_{\alpha}(a) = a^{\tfrac{\alpha}{2}-1} \;, \quad \alpha = 0, ..., N_\mathrm{B}-1\;,
\label{eq:EdS}
\end{align}
that are recursively orthonormalised over the scale factor interval $\left[ a_\mathrm{min}, 1\right]$ by Gram-Schmidt orthonormalisation to obtain the $\phi_\alpha(a)$ introduced in Equation~\eqref{eq:series_expansion}.
This basis is designed to recover the luminosity distances in an Einstein-de-Sitter universe ($\Omega_{m0} = 1$, $\Omega_\Lambda = 0$) with the first two basis functions \comm{(see \cite{bib:Mignone} for the derivation)}. It is thus a sparse basis decomposition in the matter-dominated era of the cosmic evolution, for scale factors approximately ranging from $a \approx 10^{-3}$ to $a\approx 0.7$.

\subsubsection{Variations of the Einstein-de-Sitter basis}
\label{sec:VEdS}

In the late universe ($a \gtrsim 0.7$), the term of the cosmological constant, $\Omega_\Lambda$, dominates in Equation~\eqref{eq:H_sm} and the universe expands exponentially. Determining the luminosity distances \comm{in such a cosmology}
\begin{equation}
D_\mathrm{L}(a) = \dfrac{c}{H_0} \dfrac{1}{a} \int \limits_a^1 \dfrac{\mathrm{d}x}{x^2 \sqrt{\Omega_\Lambda}}
\end{equation}
yields $D_\mathrm{L}(a) \propto 1/a^2 - 1/a$, such that a basis obtained from
\begin{align}
u_{\alpha}(a) = a^{\alpha-2} \;, \quad \alpha = 0, ..., N_\mathrm{B}-1
\label{eq:LB}
\end{align}
is physically motivated analogously to the one obtained from Equation~\eqref{eq:EdS}.

Since Equation~\eqref{eq:LB} does not include the square-root functions contained in Equation~\eqref{eq:EdS}, a combination of both bases can be obtained from functions of the form
\begin{align}
u_{\alpha}(a) = a^{\tfrac{\alpha}{2}-2} \;, \quad \alpha = 0, ..., N_\mathrm{B}-1\;.
\label{eq:comb}
\end{align}

For the reconstruction of $H(a)$, it turns out that the term in the last bracket of Equation~\eqref{eq:E} belonging to $c_0$ cancels out when using the Einstein-de-Sitter basis. This implies that the expansion function is effectively reconstructed by $N_\mathrm{B}-1$ basis functions and coefficients, such that we also set up a modified Einstein-de-Sitter basis without the first basis function from functions of the form
\begin{align}
u_{\alpha}(a) = a^{\tfrac{\alpha-1}{2}} \;, \quad \alpha = 0, ..., N_\mathrm{B}-1\;.
\label{eq:mod}
\end{align}

Table~\ref{tab:ONBs} summarises the functions that form the basis sets of Sections~\ref{sec:EdS} and \ref{sec:VEdS} after Gram-Schmidt orthonormalisation and also lists the first four functions that are employed to set up the bases. 
\comm{For a flat $\Lambda$CDM model in the late universe, 
\begin{equation}
D_\mathrm{L}(a) = \dfrac{c}{H_0} \dfrac{1}{a} \int \limits_a^1 \dfrac{\mathrm{d}x}{x^2 \sqrt{\Omega_\Lambda + \Omega_{\mathrm{m}0}x^{-3}}} \;.
\label{eq:DL_LCDM}
\end{equation}
The integral on the right-hand side cannot be simplified to a few polynomial terms as in the previous cases. Consequently, expanding the luminosity distance in such a cosmology requires an increasing amount of polynomial basis functions to achieve a highly accurate approximation. The quality of approximation for different $N_\mathrm{B}$ is investigated in Section~\ref{sec:accuracy}.}

\begin{table}
\caption{Parameter-free, analytic, orthonormal basis function sets used in this work.}
\begin{tabular}{ccccccc}
\hline 
Basis & Name & $u_\alpha(a)$ & $u_0(a)$ & $u_1(a)$ & $u_2(a)$ & $u_3(a)$ \\
\hline
\noalign{\smallskip}
1 & EdS & $a^{\alpha/2-1}$ & $1/a$ & $1/\sqrt{a}$ & $1$ & $\sqrt{a}$ \\
\noalign{\smallskip}
2 & $\Lambda$ & $a^{\alpha-2}$ & $1/a^2$ & $1/a$ & $1$ & $a$ \\
\noalign{\smallskip}
3 & comp & $a^{\alpha/2-2}$ & $1/a^2$ & $1/\sqrt{a^3}$ & $1/a$ & $1/\sqrt{a}$ \\
\noalign{\smallskip}
4 & mod & $a^{(\alpha-1)/2}$ & $1/\sqrt{a}$ & $1$ & $\sqrt{a}$ & $a$ \\
\noalign{\smallskip}
\hline
\end{tabular}
\label{tab:ONBs}
\end{table}

\subsection{Further numerical bases}

Even sparser bases that require fewer coefficients for the reconstruction exist, e.g. the principal component basis as introduced in \cite{bib:Ishida} or \cite{bib:Mignone2}. Yet, these approaches are less suitable for our purpose than an analytic basis. The latter assumes a physically motivated decomposition into basis functions that can be efficiently determined at any arbitrary point between $\left[ a_\mathrm{min}, a_\mathrm{max}\right]$, while the numerical bases require to be represented by a lot of sampling points. The dense sampling increases the run-time. It also requires numerical imprecisions to be carefully taken into account and an interpolation between the sampling points to be defined (see Section~\ref{sec:accuracy} for a comparison of a numerical and the analytic implementation of the Einstein-de-Sitter basis). 
Apart from requiring a high accuracy and precision in the lensing distance ratio, the increasing amount of data in a data set with an increasing interval of scale factors also requires a fast reconstruction of $H(a)$ and the distance measures. This favours closed form expressions over numerical ones, especially when the confidence bounds are determined by a \com{Monte-Carlo simulation} as detailed in Section~\ref{sec:confidence_bounds}.

\subsection{Confidence bounds}
\label{sec:confidence_bounds}

The uncertainties on $\hat{\boldsymbol{c}}$ due to the covariances of the $D_{\mathrm{L},i}$ are determined by a \com{Monte-Carlo simulation} from the $\chi^2$ in Equation~\eqref{eq:chi2_red}: We simulate 1000 data sets with $N_\mathrm{D}$ supernovae, each at the same scale factors as are listed in the original data set by drawing $D_{\mathrm{L},i}$, $i=1,...,N_\mathrm{D}$ from a Gaussian distribution around the measured $D_{\mathrm{L},i}$ with a width $\sigma$ corresponding to the measured uncertainty as listed in the original data set. 
\comme{In this choice of error model, we assume an ideal standardisation process, such that the $D_{\mathrm{L},i}$ determined by the light-curve fits are only subject to statistical uncertainties and are not subject to any bias anymore and that the true $D_{\mathrm{L},i}$ are equal to the reconstructed $D_{\mathrm{L},i}$ up to a linear transformation. Other choices of error models are based on drawing Gaussian random samples around a true $D_{\mathrm{L},i}$ of a simulated data set. In any case, the choice of an error model heavily relies on assumptions that are very difficult to corroborate in practice.}

Correlations between different data points are neglected because we focus on the imprecision that is caused by the measurement uncertainties in the $D_{\mathrm{L},i}$. The correlations that arise due to the compilation of several inhomogeneous data sets to one are usually much smaller than the statistical uncertainties. For the Pantheon sample, we find that the correlations, i.e. the off-diagonal entries, in $\Sigma_\mu$ are all smaller than 1\% of the statistical uncertainties.

For each of the 1000 simulated data sets, $E(a)$, $D_\mathrm{L}(a)$, and $D(z_\mathrm{l}, z_\mathrm{s})$ are reconstructed. From the ensemble of all 1000 reconstructions of each quantity, the 68\%, 95\% and 99\% confidence intervals, corresponding to 1-$\sigma$, 2-$\sigma$, and 3-$\sigma$ confidence intervals for a Gauss distribution are calculated. In addition, we determine the standard deviation of each quantity from the 1000 simulated data sets. 

While the confidence bounds on $\hat{\boldsymbol{c}}$ can also be derived from the Fisher-matrix method, the Monte-Carlo simulation is required to propagate the confidence bounds on $\hat{\boldsymbol{c}}$ accurately into $E(a)$, $D_\mathrm{L}(a)$, and $D(z_\mathrm{l},\mathrm{s})$ as these reconstructions contain non-linear transformations of $\hat{\boldsymbol{c}}$. By construction, the Monte-Carlo simulation yields the same confidence bounds on $\hat{\boldsymbol{c}}$ as can be derived from the Fisher-matrix, i.e. it reaches the Cramér-Rao lower bound as detailed in Appendix~\ref{app:Cramer_Rao}.

\subsection{Quality assessment of a basis}
\label{sec:quality_assessment}

A priori, a Friedmann-parameter-free reconstruction of $E(a)$, $D_\mathrm{L}(a)$, and $D(z_\mathrm{l}, z_\mathrm{s})$ can be performed with any set of orthonormal basis functions. To rank and compare different bases with different number of basis functions, $N_\mathrm{B}$, we assess their reconstruction quality \comm{by calculating the reduced $\chi^2$ defined as
\begin{align}
\chi_\nu^2 \equiv \dfrac{\chi^2}{N_\mathrm{D}-N_\mathrm{B}} \;,
\label{eq:red_chi2}
\end{align}
and we determine the relative imprecision of the reconstruction given by the confidence bounds determined according to Section~\ref{sec:confidence_bounds}.}

The optimal reconstruction has $\chi_\nu^2 = 1$. If $\chi_\nu^2  > 1$, the basis does not fully capture the information contained in the data or the covariances have been underestimated. For $\chi_\nu^2 < 1$, the basis overfits the data or the covariances have been overestimated. \comm{Hence, monitoring $\chi^2_\nu(N_\mathrm{B})$ for an increasing number of basis functions, we can determine the basis with maximum information about $E(a)$, $D_\mathrm{L}(a)$, and $D(z_\mathrm{l}, z_\mathrm{s})$, i.e.\@ which has the maximum number of $N_\mathrm{B}$ retrievable for a given measurement precision and which fulfils $\chi^2_\nu(N_\mathrm{B}) \approx 1$. Bases with small confidence bounds and small $N_\mathrm{B}$ are preferred.}

\comm{In addition, we determine the reconstruction accuracy of $E(a)$, $D_\mathrm{L}(a)$, and $D(z_\mathrm{l}, z_\mathrm{s})$ for a basis in a simulation of a flat $\Lambda$CDM model. Since the Pantheon sample can be explained by such a cosmological model, the result is a good estimate for the reconstruction accuracy of the true, unknown cosmology of our universe. The reconstruction inaccuracies of $E(a)$ in the $\Lambda$CDM model simulation should lie within the confidence bounds of $E(a)$ caused by the measurement precision. Assuming that the inaccuracy of the data-driven reconstruction of $E(a)$ in the true cosmology is of the same order of magnitude as the inaccuracy in the $\Lambda$CDM model simulation, we do not expect the data-driven reconstruction to be biased.}

To summarise, we search for the basis which comes closest to $\chi_\nu^2 =1$ \comm{for the maximum number of $N_\mathrm{B}$ retrievable for a given measurement precision}, shows the smallest bias \comm{in a $\Lambda$CDM model simulation} and has the smallest confidence bounds (i.e. relative imprecisions) that encompass the relative inaccuracies \comm{of the $\Lambda$CDM model simulation} at the 68\% confidence level.

\subsection{Run-time enhancements}
\label{sec:computational_details}

Our implementation is based on MATLAB, employing the full covariance matrix including the correlations between the data points.
As detailed in Section~\ref{sec:basis_selection}, we choose the Einstein-de-Sitter basis as orthonormal set of basis functions. Analytically performing the Gram-Schmidt orthonormalisation, we obtain closed-form expressions for $\Phi$ and subsequently closed-form expressions for $D_\mathrm{L}(a,\hat{\boldsymbol{c}})$, $H(a)$, $D_\mathrm{L}(a)$, and $D(z_\mathrm{l},z_\mathrm{s})$ up to $N_\mathrm{B}=4$. Higher order coefficients and basis functions can be retrieved numerically. Yet, for the Pantheon sample, maximally four coefficients are significantly determined given the covariances (see Section~\ref{sec:basis_selection}).

A major speed-up in run-time is obtained by replacing the standard \emph{inv{}}-function by \emph{mldivide()} to calculate $\hat{\boldsymbol{c}}$ and by replacing all \emph{for}-loops by matrix operations. 
The overall run-time to reconstruct $H(a)$, $D_\mathrm{L}(a)$, and $D(z_\mathrm{l},z_\mathrm{s})$ (the latter for one fixed $z_\mathrm{l}$, as detailed in Section~\ref{sec:d}) including confidence bounds according to Section~\ref{sec:confidence_bounds} for the Pantheon sample amounts to roughly 110 seconds on a standard notebook (MacBook Pro, 2.2 GHz Intel Core i7, 8 GB 1333 MHz DDR3 RAM). 


\section{Application to data}
\label{sec:application}

\subsection{Synopsis of the data}
\label{sec:data}

As observational data, we use the Pantheon data set, \cite{bib:Scolnic}, for the selection of the optimum basis function (Section~\ref{sec:basis_selection}) and to determine the reconstruction precision for $E(a)$, $D_\mathrm{L}(a)$, and $D(z_\mathrm{l},z_\mathrm{s})$ (Section~\ref{sec:precision}). To investigate the reconstruction accuracies \comm{for different implementations and for varying $N_\mathrm{B}$} (Section~\ref{sec:accuracy}), we generate Pantheon-like simulations. We simulate luminosity distances at the scale factors of the Pantheon sample from a $\Lambda$CDM model as parametrised by \cite{bib:Planck}, based on measurements from the cosmic microwave background, (see first column of Table~\ref{tab:reference_LCDM}) and divide them by an arbitrary scale to obtain scale-free $\tilde{D}_{\mathrm{L},i}$. The parametrisation in the second row of Table~\ref{tab:reference_LCDM} by \cite{bib:Scolnic} is based on \com{a fit to} the Pantheon sample and measurements of $H_0$ in the local neighbourhood \com{(employing $H_0$ from \cite{bib:Riess_H0})}. \com{We use these two parametrised $\Lambda$CDM models in Section~\ref{sec:precision} and in Section~\ref{sec:model_based_reconstructions} for the comparison with our reconstructions.}

\begin{table}
\caption{$\Lambda$CDM parametrisations of \protect\cite{bib:Planck} (first row) and \protect\cite{bib:Scolnic} (second row).}
\begin{tabular}{cccccc}
\hline 
$\Lambda$CDM & $\Omega_\mathrm{r0}$ & $\Omega_\mathrm{m0}$ & $\Omega_\mathrm{K}$ & $\Omega_{\Lambda}$ & $H_0$ \\
model & & & & & [km/s/Mpc] \\
\hline
(Planck) & 0.0 & 0.3089 & 0.0 & 0.6911 & 67.74 \\
(Scolnic) & 0.0 & 0.298 & 0.0 & 0.702 & 73.52$^{(1)}$ \\
\hline
\end{tabular}
$^{(1)}$taken from \citet{bib:Riess_H0}
\label{tab:reference_LCDM}
\end{table}
\subsection{Selection of the optimal basis}
\label{sec:basis_selection}

\begin{figure*}
\centering
\includegraphics[width=0.32\textwidth]{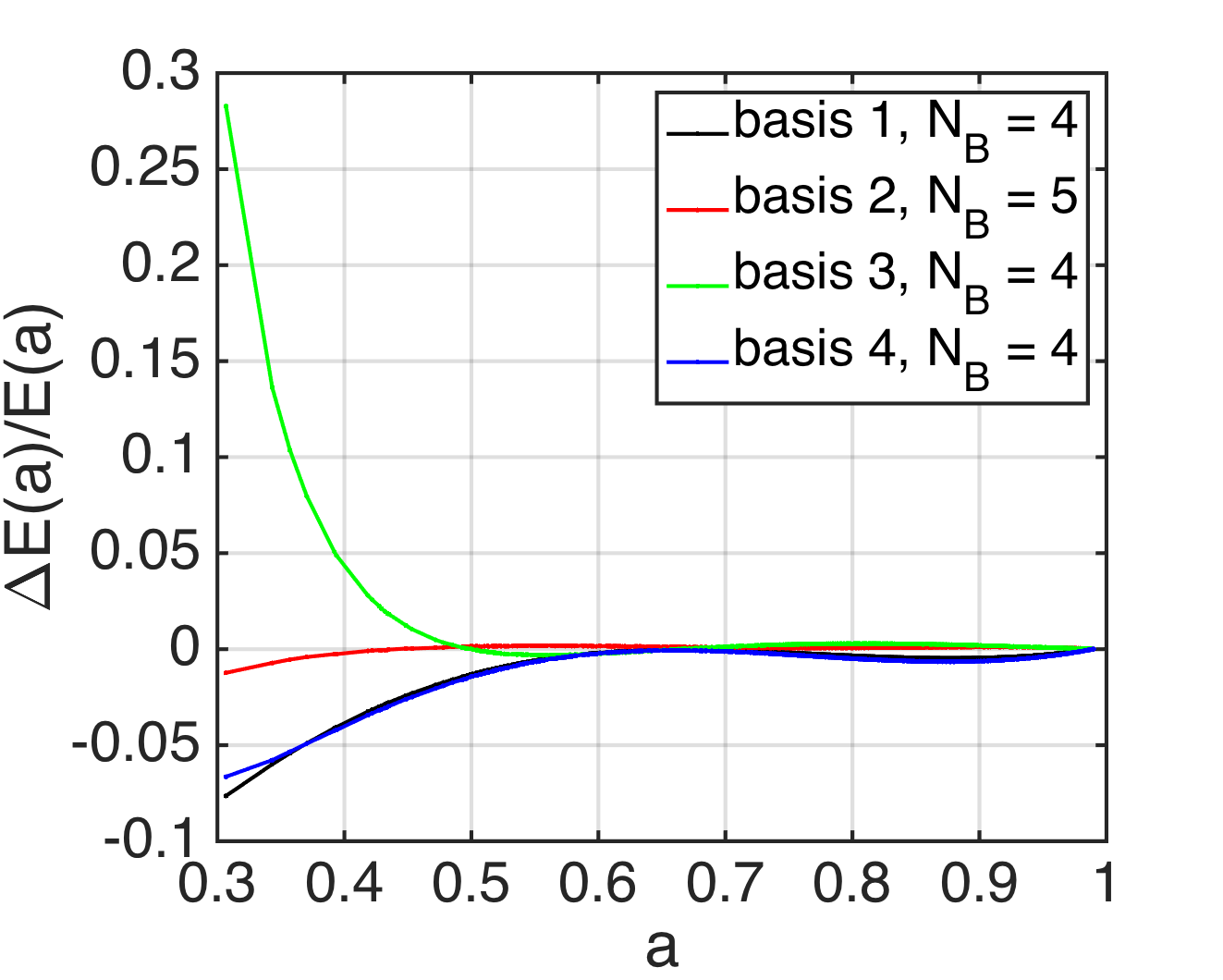} 
\includegraphics[width=0.32\textwidth]{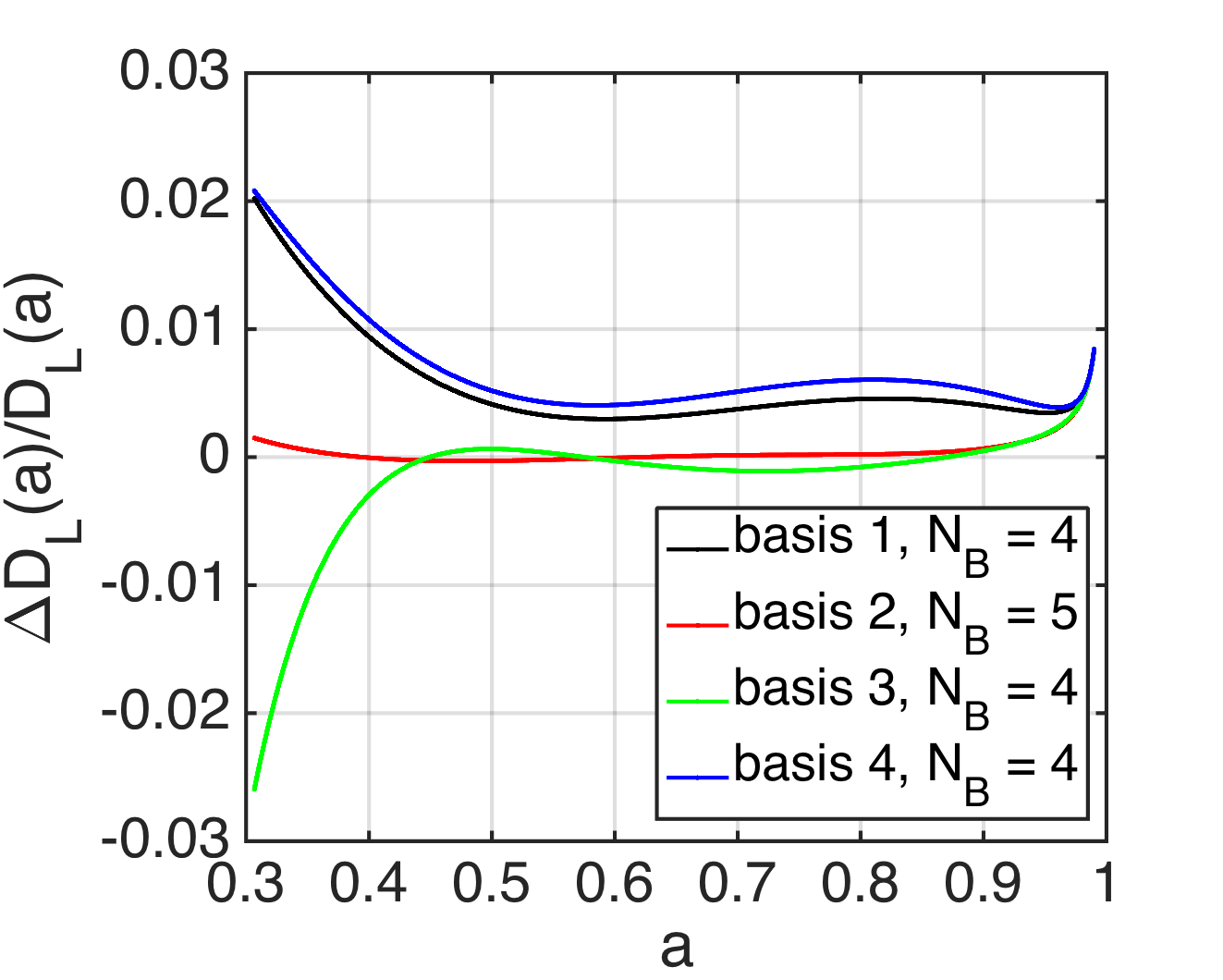}
\includegraphics[width=0.32\textwidth]{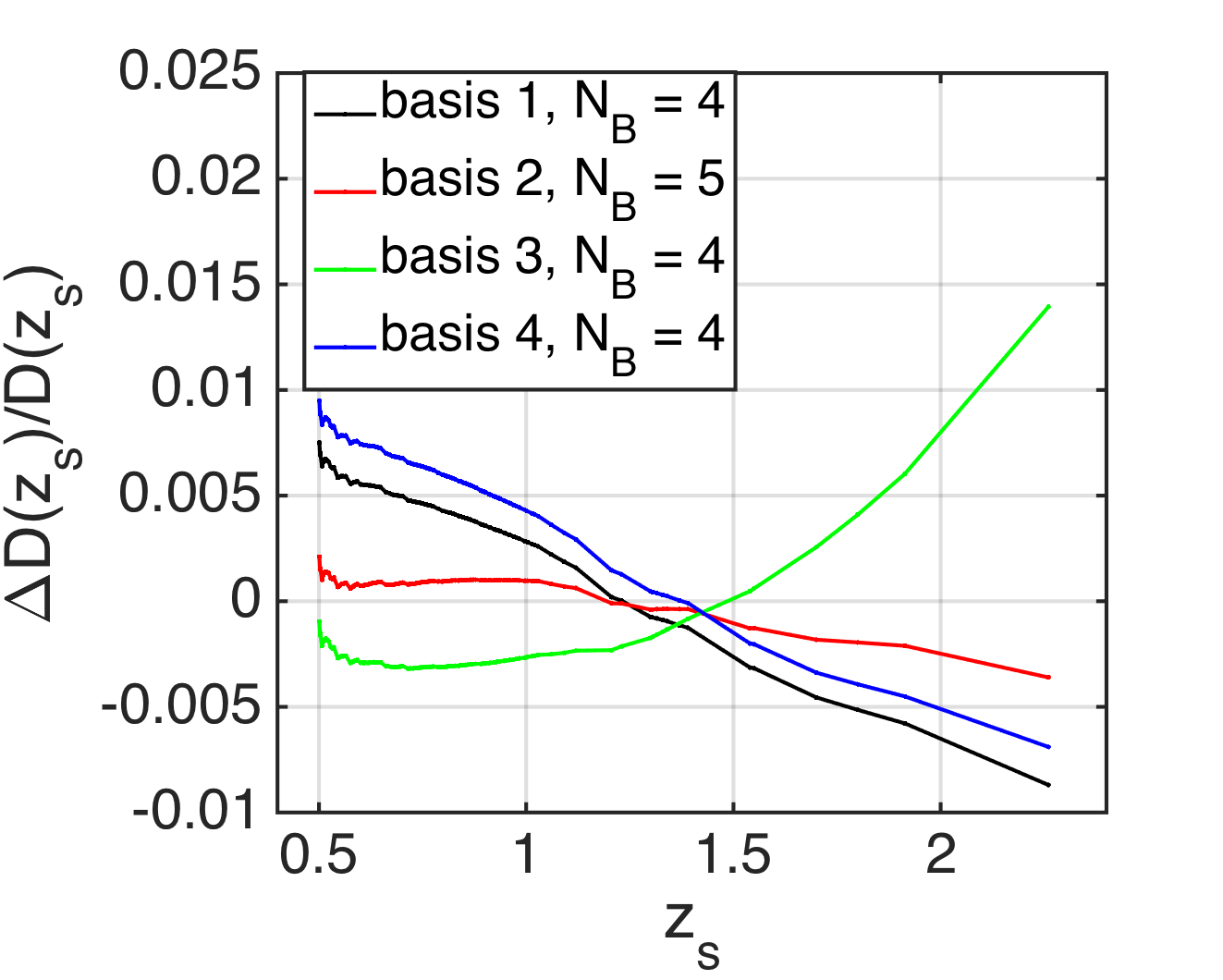}
\includegraphics[width=0.32\textwidth]{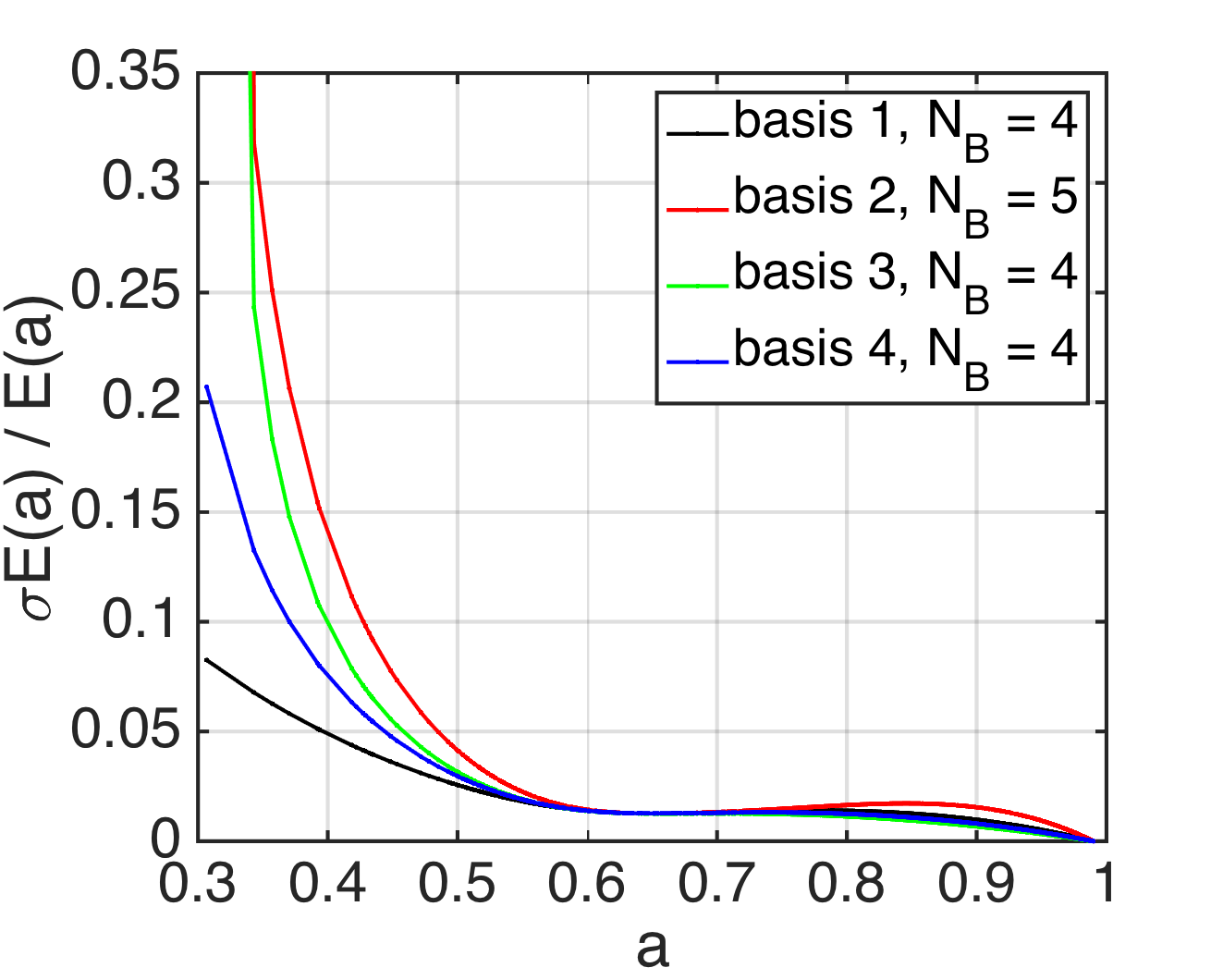} 
\includegraphics[width=0.32\textwidth]{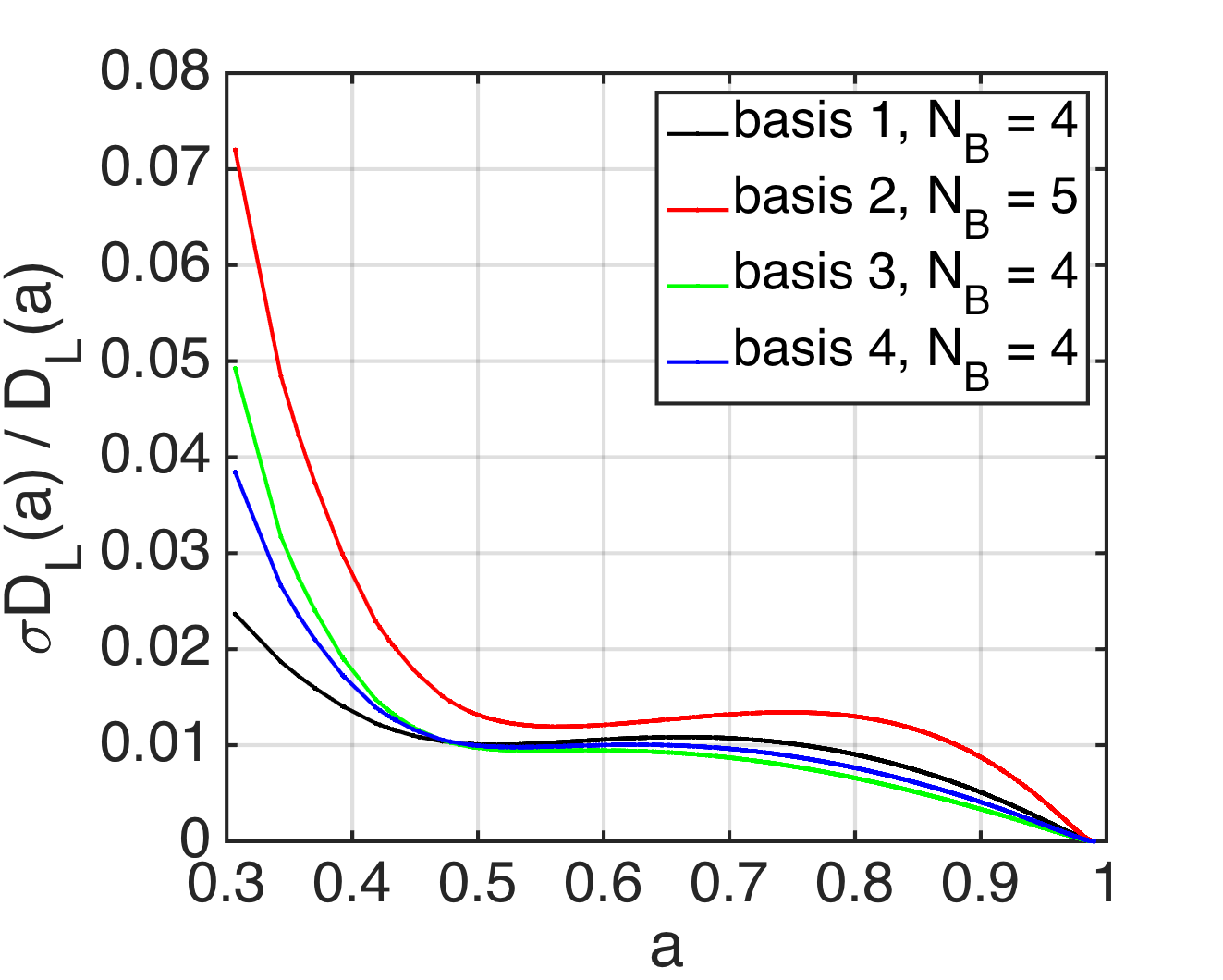}
\includegraphics[width=0.32\textwidth]{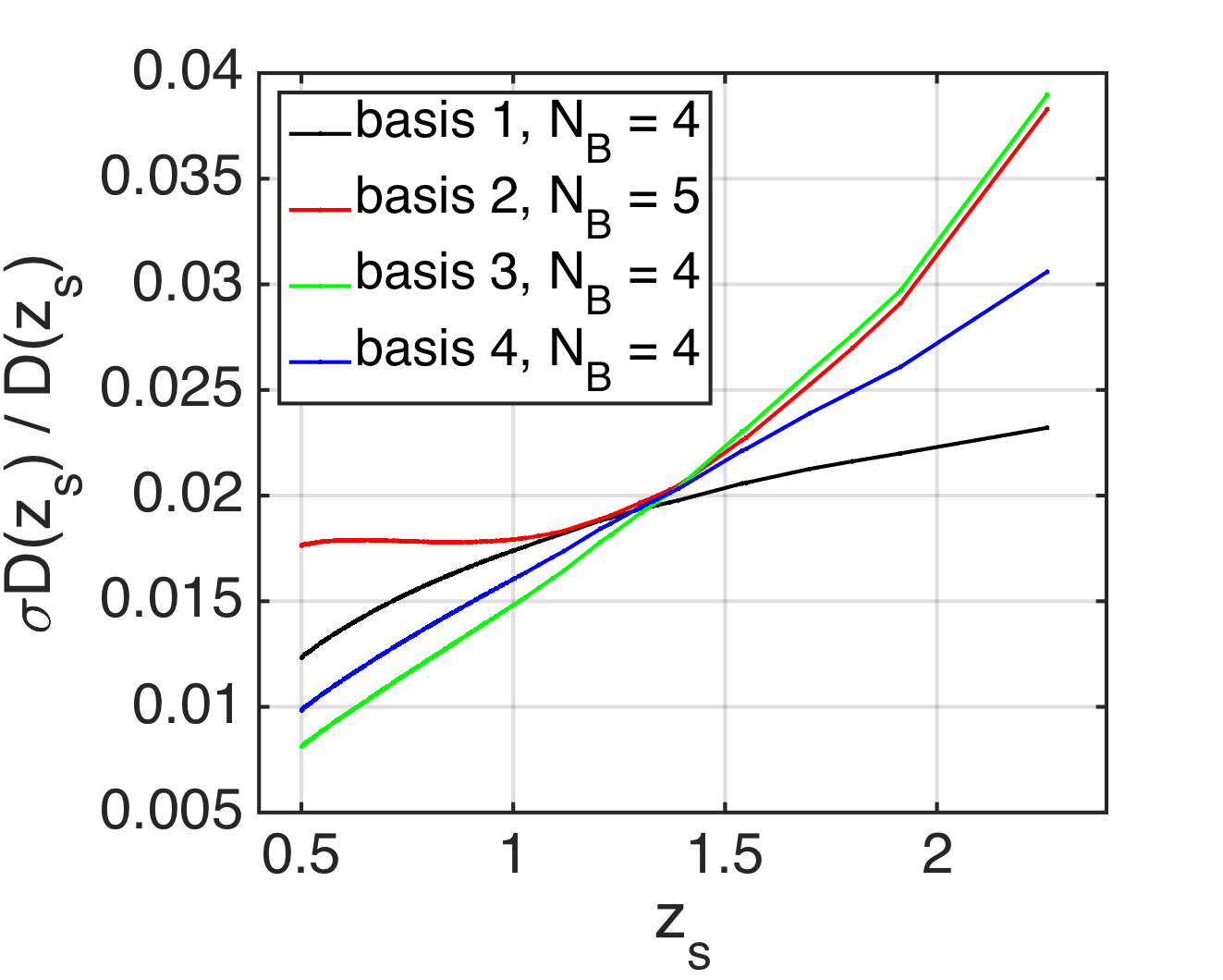}
\caption{Top (from left to right): Relative inaccuracy of $E(a)$, $D_\mathrm{L}(a)$, and $D(0.5,z_\mathrm{s})$ w.r.t. the ones from a $\Lambda$CDM model as parametrised in \protect\cite{bib:Planck}. Bottom (from left to right): Relative imprecisions (standard deviations of the \com{Monte-Carlo simulation based on the Pantheon data set}) of $E(a)$, $D_\mathrm{L}(a)$, and $D(0.5,z_\mathrm{s})$ for the same basis configurations.}
\label{fig:basis_selection}
\end{figure*}

We first perform a comparison of the bases listed in Table~\ref{tab:ONBs} to find the optimal basis and number of basis functions, $N_\mathrm{B}$, given the quality assessment criteria of Section~\ref{sec:quality_assessment}. Table~\ref{tab:basis_selection} shows the quality measures for different configurations of the basis sets. For all bases 1 to 4, we test 2 to 6 numerically implemented basis functions. \comm{For the numerical implementation, we equidistantly sub-sample $a$} at $N_\mathrm{S}= 10 000$ points between $a_\mathrm{min}$ and 1. We employ the full covariance matrix in Equation~\eqref{eq:chi2_red} and determine confidence bounds as detailed in Section~\ref{sec:confidence_bounds} from the data of the Pantheon sample.

Comparing $\hat{\boldsymbol{c}}$ for basis 1 for all $N_\mathrm{B}$ with the $\hat{\boldsymbol{c}}$ obtained when using the diagonal covariance matrix with the statistical uncertainties only, we find that the coefficients deviate only on the order of $10^{-4}$. Hence, correlations between the different data in the Pantheon sample play a minor role for the reconstructions of $E(a)$, $D_\mathrm{L}(a)$, and $D(z_\mathrm{l},z_\mathrm{s})$\comm{, as already mentioned in Section~\ref{sec:taylor_series}}.

For all bases, $N_\mathrm{B} >3$ yields $\chi^2_\nu$ slightly smaller than one, as listed in the third column of Table~\ref{tab:basis_selection}, so that we conclude that these basis configurations capture the information contained in the data well. Contrary to that $N_\mathrm{B}=2$ has  $\chi^2_\nu > 1$ for all bases, indicating that these configurations are not suitable to represent $D_\mathrm{L}(a)$. 

Next, we determine the expected inaccuracies from the simulated Pantheon-like data set \comm{in a flat $\Lambda$CDM model simulation}, as detailed in Section~\ref{sec:data}. We observe that the relative inaccuracies are contained within the 68\% confidence bounds for configurations of $N_\mathrm{B}>3$ for bases 1,3, and 4 and for $N_\mathrm{B}>5$ for basis 2. For these configurations, we compare the standard deviations of $\hat{\boldsymbol{c}}$ \comm{with $\hat{\boldsymbol{c}}$} for all $N_\mathrm{B}$ to find the best basis configuration for each of the four basis sets as follows: 
\begin{itemize}
\item basis 1 with $N_\mathrm{B} = 4$, 
\item basis 2 with $N_\mathrm{B} = 5$, 
\item basis 3 with $N_\mathrm{B} = 4$,
\item basis 4 with $N_\mathrm{B} = 4$.
\end{itemize}

\comm{For basis~2, more than 6 significant coefficients could be determined. Yet, monitoring the continuously declining $\chi^2_\nu(N_\mathrm{B})$ for this basis and the increasing width of the confidence bounds, this basis is practically of much less use than the others, so that we consider the version of $N_\mathrm{B}=5$ for the sake of completeness in the following.}

\begin{sidewaystable*}
\caption{Quality assessment of the bases of Table~\ref{tab:ONBs} according to Section~\ref{sec:quality_assessment}.}
\label{tab:basis_selection}
\begin{tabular}{cccccccccccccccc}
\hline 
\noalign{\smallskip}
B & $N_\mathrm{B}$ & $\chi_\nu^2$ & P>A & $\hat{c}_0$ & $\hat{c}_1$ & $\hat{c}_2$ & $\hat{c}_3$ & $\hat{c}_4$ & $\hat{c}_5$ & $\sigma_{c_0}$ & $\sigma_{c_1}$ & $\sigma_{c_2}$ & $\sigma_{c_3}$ & $\sigma_{c_4}$ & $\sigma_{c_5}$\\
\noalign{\smallskip}
\hline
\noalign{\smallskip}
1	&	2	&	1.5534	&	\xmark	&	0.5658	&	-0.2402	&		&		&		&		&	0.0013	&	0.0006	&		&		&		&		\\
1	&	3	&	0.9065	&	\xmark	&	0.6805	&	-0.3229	&	0.0264	&		&		&		&	0.0045	&	0.0032	&	0.0010	&		&		&		\\
1	&	4	&	0.9063	&	\cmark	&	0.6717	&	-0.3158	&	0.0217	&	0.0013	&		&		&	0.0094	&	0.0083	&	0.0046	&	0.0012	&		&		\\
1	&	5	&	0.9069	&	\cmark	&	0.6649	&	-0.3087	&	0.0162	&	0.0041	&	-0.0007	&		&	0.0146	&	0.0144	&	0.0102	&	0.0048	&	0.0012	&		\\
1	&	6	&	0.9052	&	\cmark	&	0.6515	&	-0.2926	&	-0.0001	&	0.0169	&	-0.0076	&	0.0019	&	0.0163	&	0.0169	&	0.0136	&	0.0087	&	0.0042	&	0.0011	\\
\noalign{\smallskip}
\hline																													\noalign{\smallskip}						
2	&	2	&	1.7080	&	\xmark	&	1.0009	&	-0.2296	&		&		&		&		&	0.0024	&	0.0006	&		&		&		&		\\
2	&	3	&	0.9439	&	\xmark	&	0.8121	&	-0.1480	&	-0.0238	&		&		&		&	0.0069	&	0.0028	&	0.0008	&		&		&		\\
2	&	4	&	0.9072	&	\xmark	&	0.7397	&	-0.1064	&	-0.0464	&	0.0061	&		&		&	0.0132	&	0.0071	&	0.0036	&	0.0010	&		&		\\
2	&	5	&	0.9070	&	\cmark	&	0.7246	&	-0.0961	&	-0.0538	&	0.0098	&	-0.0010	&		&	0.0191	&	0.0118	&	0.0077	&	0.0035	&	0.0009	&		\\
2	&	6	&	0.9050	&	\cmark	&	0.7060	&	-0.0810	&	-0.0679	&	0.0204	&	-0.0066	&	0.0015	&	0.0212	&	0.0141	&	0.0106	&	0.0066	&	0.0031	&	0.0008	\\
\noalign{\smallskip}
\hline																													\noalign{\smallskip}			
3	&	2	&	2.9099	&	\xmark	&	1.1351	&	-0.2941	&		&		&		&		&	0.0028	&	0.0008	&		&		&		&		\\
3	&	3	&	0.9144	&	\xmark	&	0.7629	&	-0.1150	&	-0.0511	&		&		&		&	0.0083	&	0.0039	&	0.0011	&		&		&		\\
3	&	4	&	0.9055	&	\cmark	&	0.7179	&	-0.0858	&	-0.0668	&	0.0041	&		&		&	0.0161	&	0.0098	&	0.0049	&	0.0013	&		&		\\
3	&	5	&	0.9064	&	\cmark	&	0.7168	&	-0.0849	&	-0.0675	&	0.0045	&	-0.0001	&		&	0.0209	&	0.0144	&	0.0094	&	0.0044	&	0.0011	&		\\
3	&	6	&	0.9064	&	\cmark	&	0.7099	&	-0.0773	&	-0.0759	&	0.0116	&	-0.0041	&	0.0011	&	0.0219	&	0.0162	&	0.0126	&	0.0083	&	0.0042	&	0.0011	\\
\noalign{\smallskip}
\hline																													\noalign{\smallskip}						
4	&	2	&	1.5841	&	\xmark	&	0.7229	&	-0.4075	&		&		&		&		&	0.0017	&	0.0010	&		&		&		&		\\
4	&	3	&	0.9510	&	\xmark	&	0.9193	&	-0.5750	&	0.0419	&		&		&		&	0.0076	&	0.0064	&	0.0016	&		&		&		\\
4	&	4	&	0.9062	&	\cmark	&	0.8163	&	-0.4693	&	-0.0050	&	0.0105	&		&		&	0.0164	&	0.0163	&	0.0068	&	0.0015	&		&		\\
4	&	5	&	0.9064	&	\cmark	&	0.8026	&	-0.4536	&	-0.0146	&	0.0148	&	-0.0010	&		&	0.0229	&	0.0246	&	0.0133	&	0.0053	&	0.0012	&		\\
4	&	6	&	0.9062	&	\cmark	&	0.7925	&	-0.4393	&	-0.0273	&	0.0242	&	-0.0060	&	0.0013	&	0.0244	&	0.0274	&	0.0171	&	0.0098	&	0.0046	&	0.0012	\\
\noalign{\smallskip}
\hline
\end{tabular}
\\
the 4th column checks whether the relative imprecision (P) at 68\% confidence level is larger than the relative inaccuracy (A) \comm{in the flat $\Lambda$CDM model simulation}, and the remainder lists the $\hat{\boldsymbol{c}}$ and its standard deviation obtained from the Monte-Carlo simulation.
\end{sidewaystable*}

Basis 1 with $N_\mathrm{B} = 4$ turns out to have the smallest confidence bounds and thus the least relative imprecision for the reconstructions of $E(a)$, $D_\mathrm{L}(a)$, and $D(z_\mathrm{l},z_\mathrm{s})$ from the Pantheon sample. Figure~\ref{fig:basis_selection} shows the relative inaccuracies of the Pantheon-like \comm{$\Lambda$CDM model} simulation and relative imprecisions for the reconstructions of the Pantheon sample in form of the standard deviations of the \com{Monte-Carlo simulation} as obtained according to Section~\ref{sec:confidence_bounds}. For the plots showing the relative inaccuracies, the model-based reconstructions from $\Lambda$CDM (\cite{bib:Planck}) are subtracted from our reconstructions \com{of the simulated Pantheon-like data set}.  

\begin{figure*}
\centering
  \includegraphics[width=0.489\textwidth]{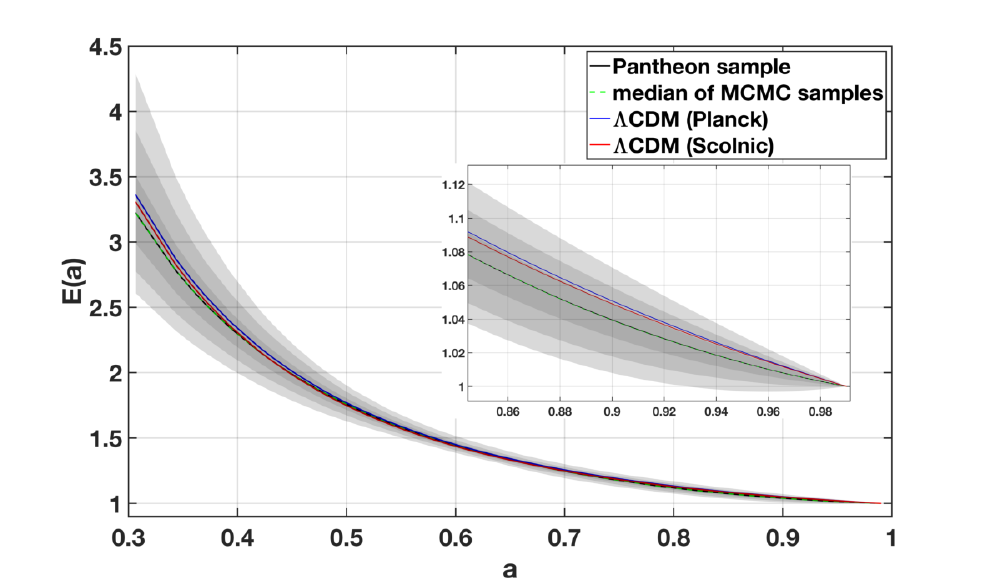} 
  \includegraphics[width=0.489\textwidth]{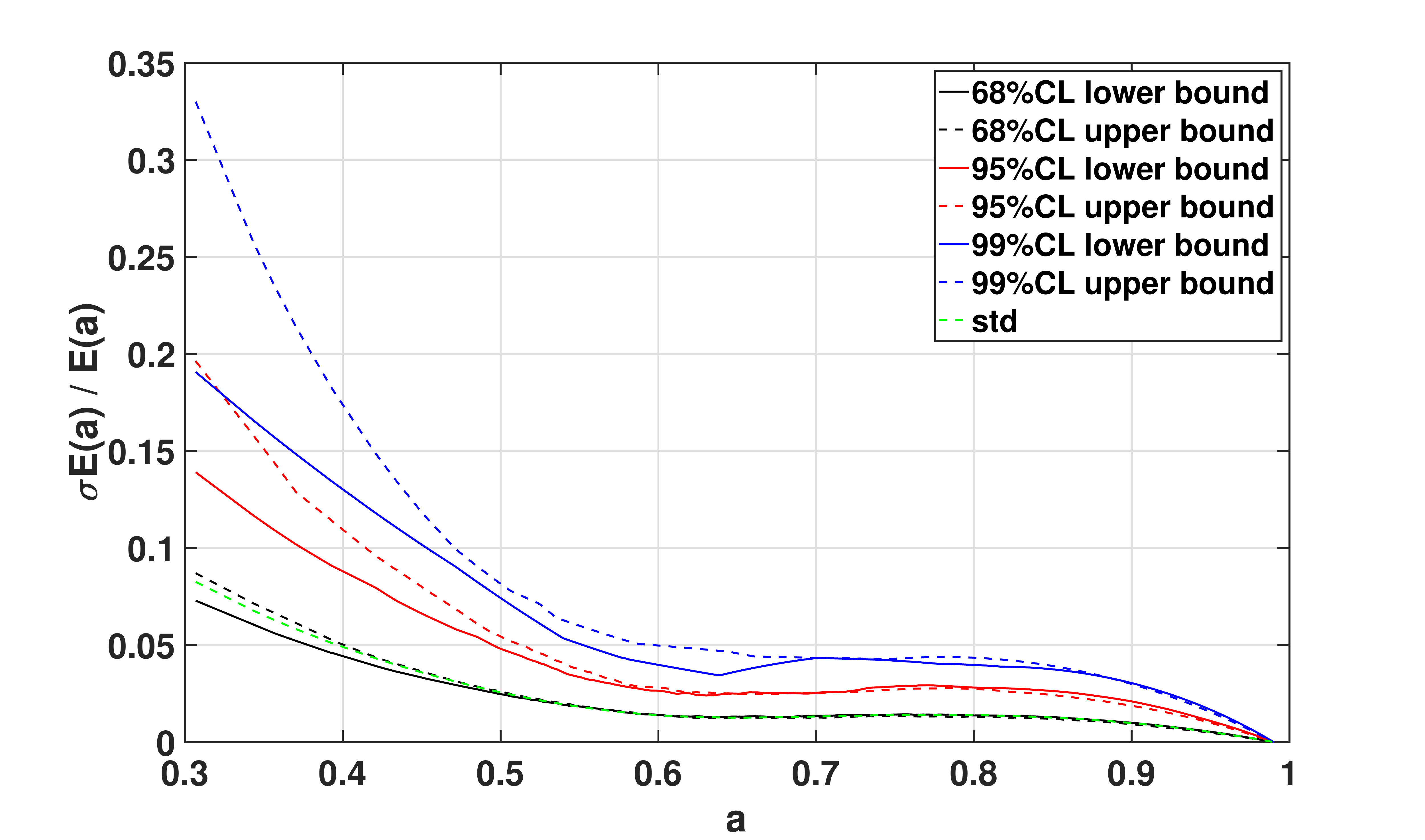} \\
 \includegraphics[width=0.489\textwidth]{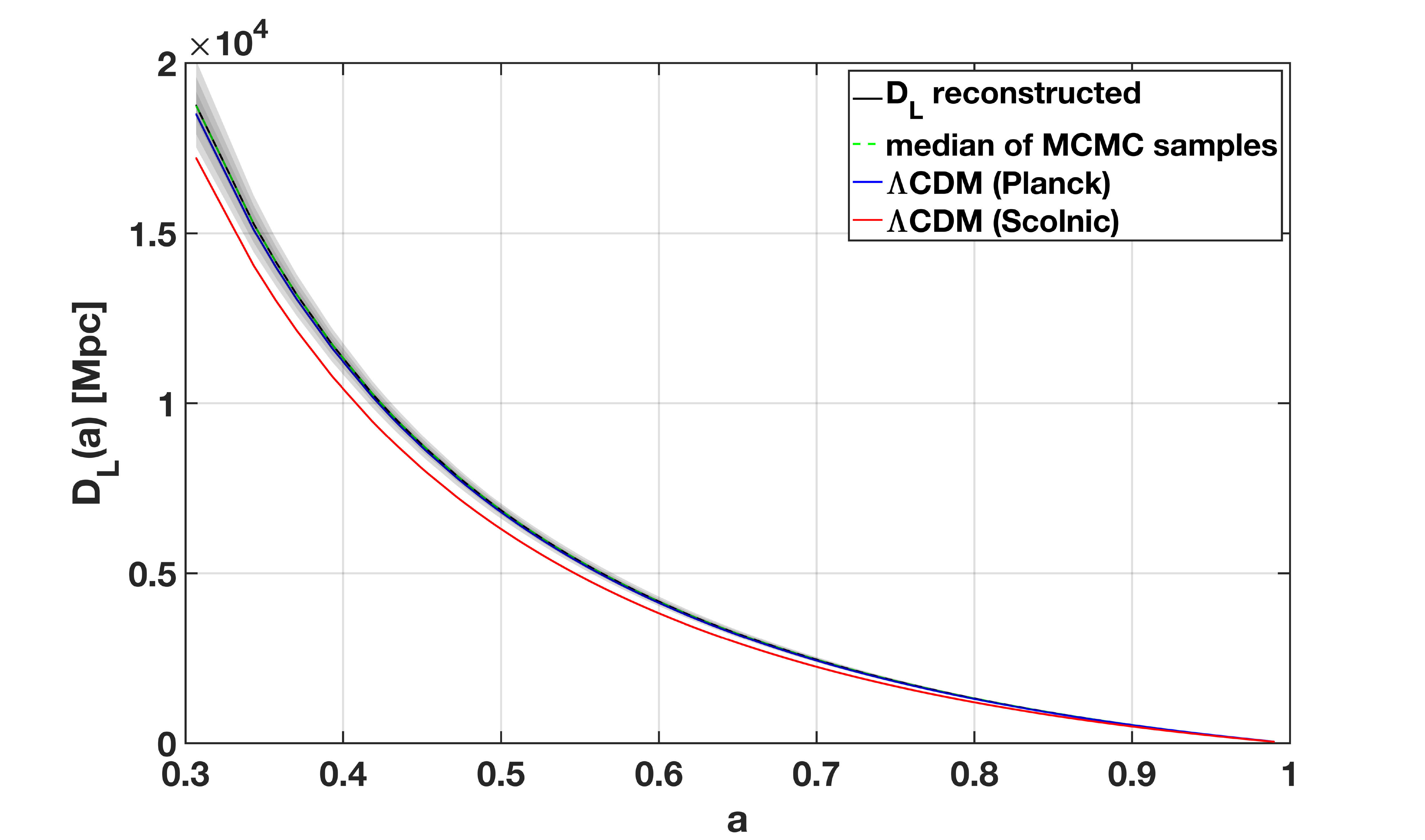} 
  \includegraphics[width=0.489\textwidth]{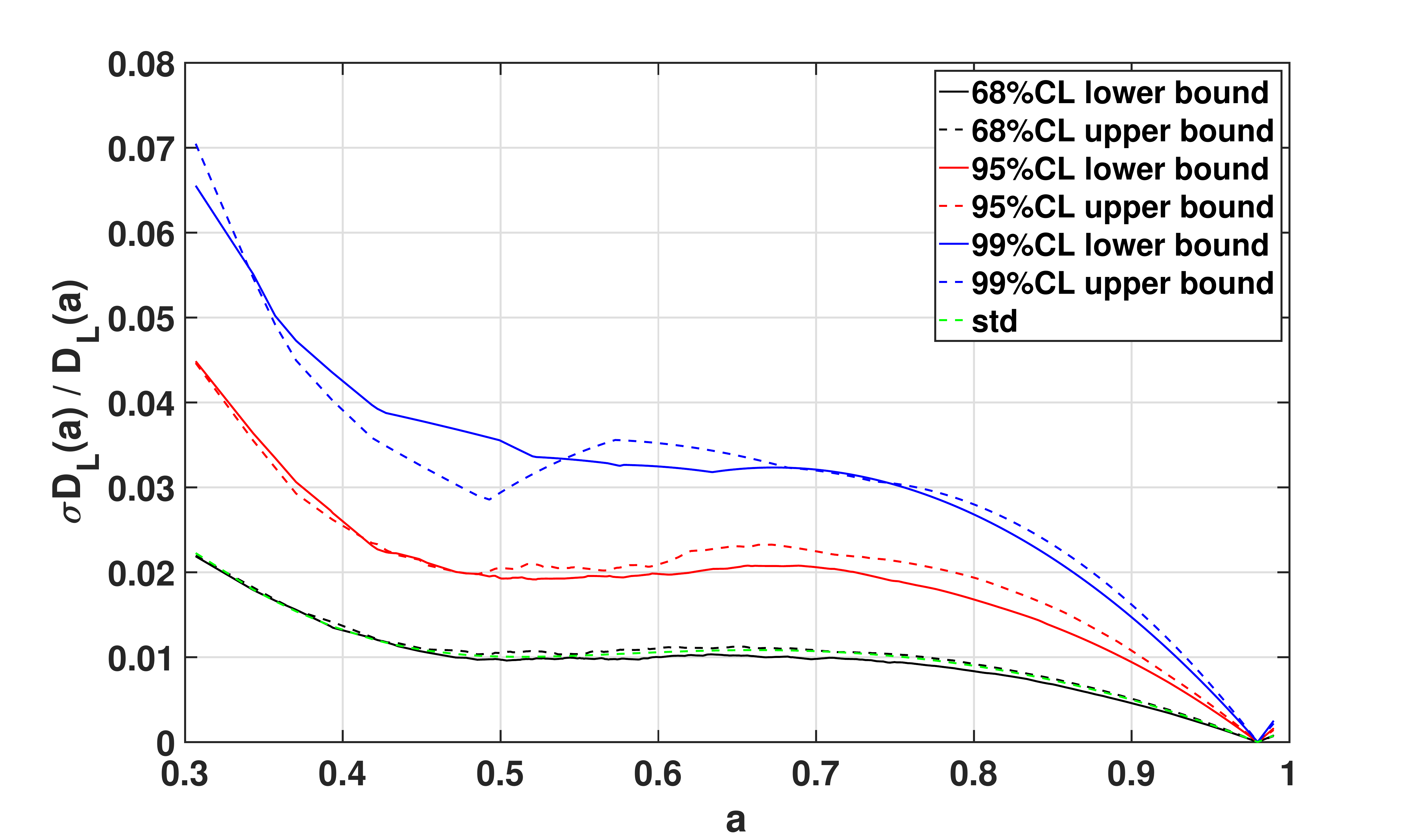} \\
 \includegraphics[width=0.489\textwidth]{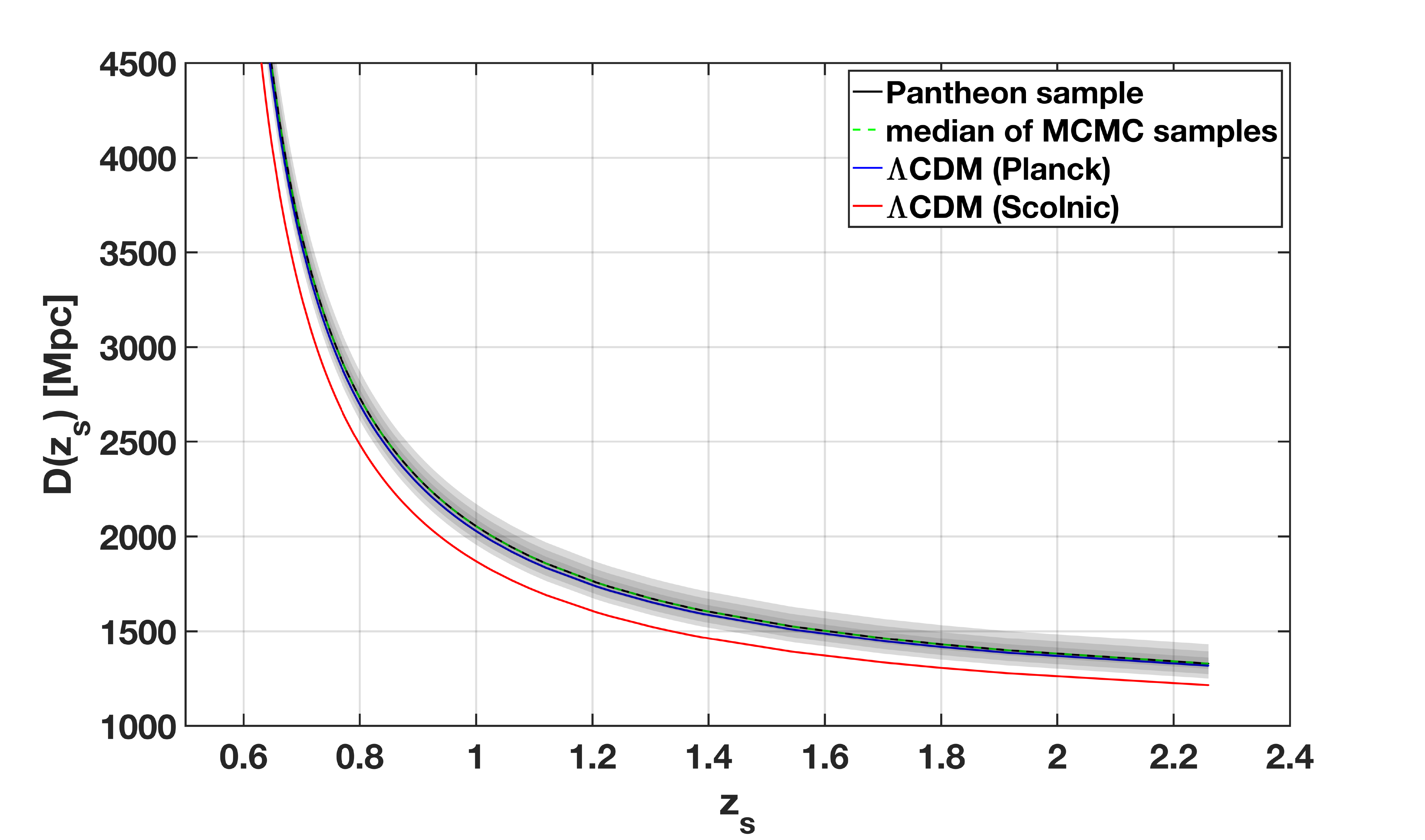}
  \includegraphics[width=0.489\textwidth]{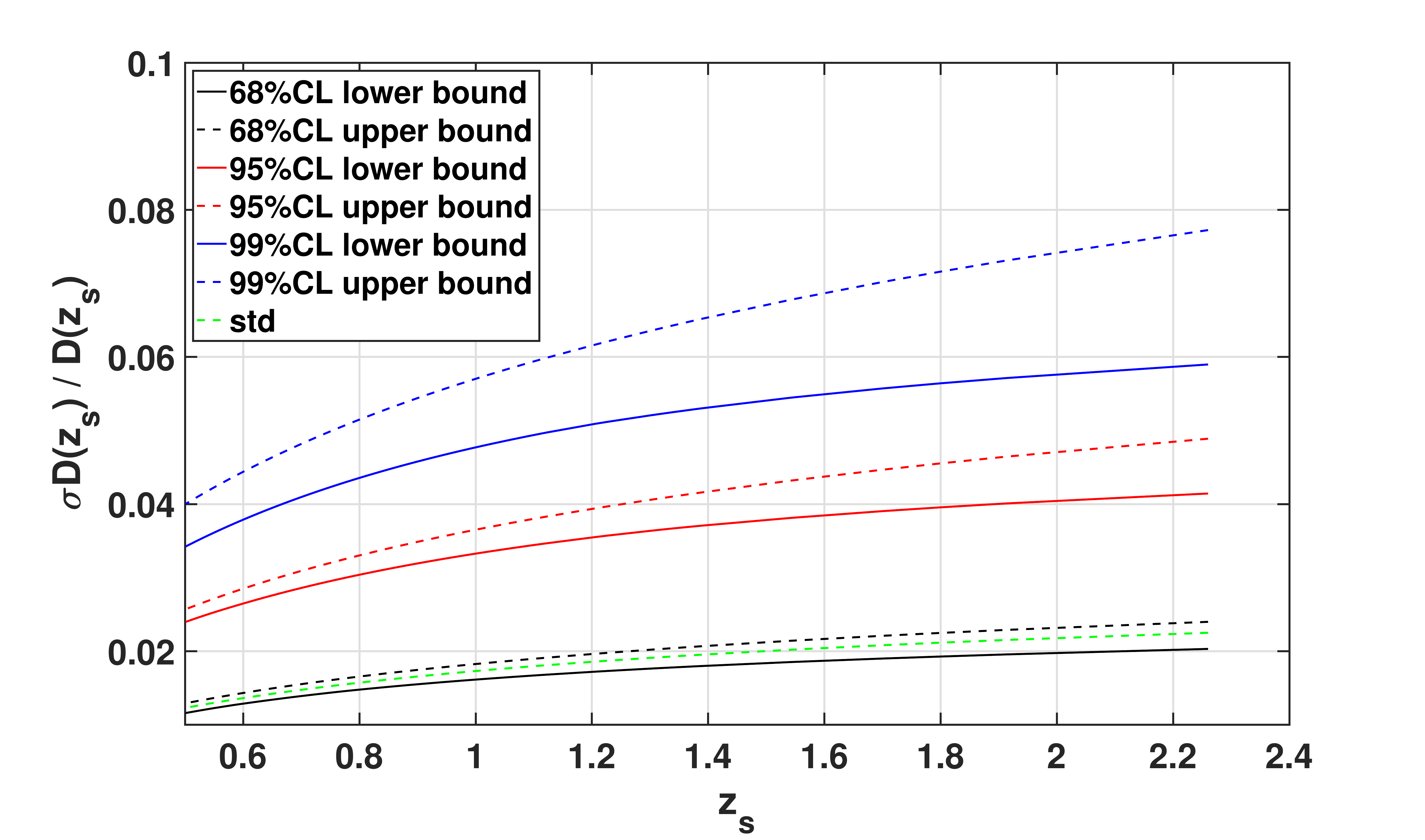}
   \caption{Top row: $E(a)$ compared to the one from a $\Lambda$CDM model as parametrised in \protect\cite{bib:Planck} and \protect\cite{bib:Scolnic} all normalised to $E(a_\mathrm{max})$ (left) and its relative reconstruction imprecision (right), Central row: $D_\mathrm{L}(a)$ compared to the one from a $\Lambda$CDM model as parametrised in \protect\cite{bib:Planck} and \protect\cite{bib:Scolnic} (left) and its relative reconstruction imprecision using the scaling of Equation~\eqref{eq:H_scaled} (right), Bottom row: $D(0.5,z_\mathrm{s})$ based on $D_\mathrm{L}(a)$ shown in the central row compared to the one from a $\Lambda$CDM model as parametrised in \protect\cite{bib:Planck} and \protect\cite{bib:Scolnic} (left) and its relative reconstruction imprecision (right). \com{We plot the relative imprecisions, as they are independent of $H_0$.}}
\label{fig:Pantheon_precision}
\end{figure*}

Thus, unless mentioned otherwise, all reconstructions in the following are determined with the standard settings for our MATLAB code as detailed in Section~\ref{sec:implementational_details}, i.e. we employ
\begin{itemize}
\item the full scale-free covariance matrix, $\tilde{\Sigma}$ from the Pantheon sample (see Equation~\eqref{eq:sigma}),
\item the scale-free Einstein-de-Sitter basis of Section~\ref{sec:EdS} with $N_\mathrm{B}=4$ basis functions (see Equation~\eqref{eq:series_expansion_red}),
\item closed-form expressions for the basis functions, their derivatives to reconstruct $E(a)$, $D_\mathrm{L}(a)$, and $D(z_\mathrm{l},z_\mathrm{s})$ (Equations~\eqref{eq:E_full}, \eqref{eq:D_L}, \eqref{eq:D}) as detailed in Section~\ref{sec:theoretical_derivations},
\item the scaling according to Equation~\eqref{eq:H_scaled} with $H(a_\mathrm{max})$ as determined by the $\Lambda$CDM model based on \cite{bib:Planck},
\item 1000 samples in a \com{Monte-Carlo simulation} drawn from the Pantheon data set to calculate confidence bounds as detailed in Section~\ref{sec:confidence_bounds}. \comm{We also use 1000 samples for the Pantheon-like simulations of flat $\Lambda$CDM models.}
\end{itemize}

\subsection{Reconstruction precision from the Pantheon sample}
\label{sec:precision}

\begin{figure*}
\centering
\includegraphics[width=0.32\textwidth]{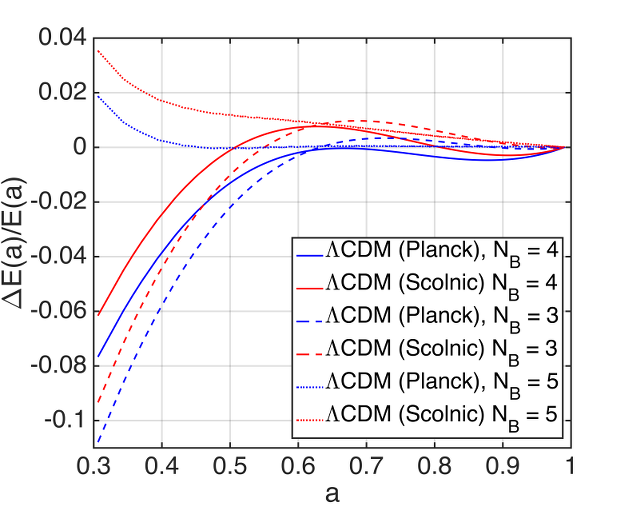} 
\includegraphics[width=0.32\textwidth]{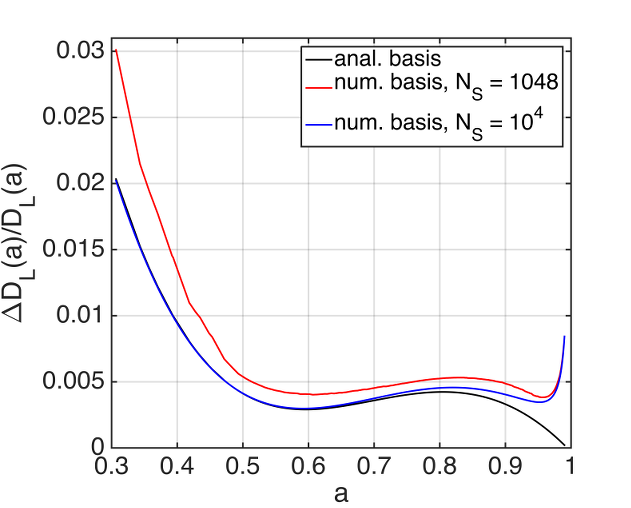}
\includegraphics[width=0.32\textwidth]{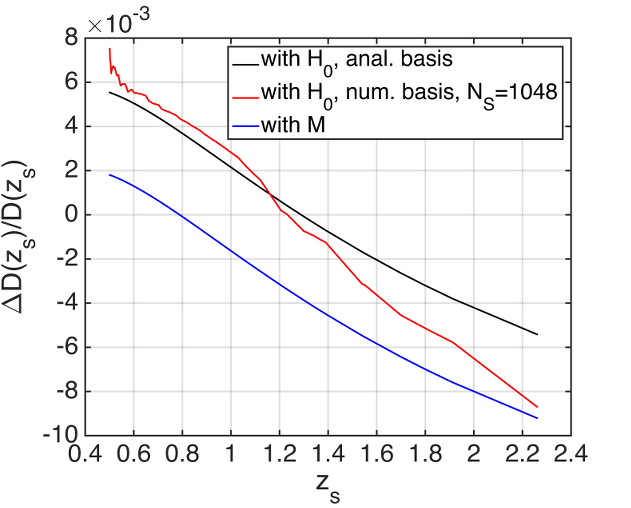}
\caption{Left: Relative inaccuracy of $E(a)$ compared to the one from a $\Lambda$CDM model as parametrised in \protect\cite{bib:Planck} and \protect\cite{bib:Scolnic} all normalised to $E(a_\mathrm{max})$ for $N_\mathrm{B}=3,4,5$ basis functions, Centre: Relative inaccuracy of $D_\mathrm{L}(a)$ compared to the one from a $\Lambda$CDM model as parametrised in \protect\cite{bib:Planck} for analytic basis functions and numerical basis functions with different number of sampling points, Right: Relative inaccuracy of $D(0.5,z_\mathrm{s})$ based on $D_\mathrm{L}(a)$ shown in the centre for a scaling with $H(a_\mathrm{max})$ according to Equation~\eqref{eq:H_scaled} and when using a scaling with $M$ according to Equation~\eqref{eq:series_expansion_red}.}
\label{fig:Pantheon_accuracy}
\end{figure*}

Running the MATLAB code \com{on the Pantheon sample} with all specifications as described at the end of Section~\ref{sec:basis_selection}, we obtain $E(a)$, $D_\mathrm{L}(a)$, and $D(0.5,z_\mathrm{s})\equiv D(z_\mathrm{s})$ as shown in Figure~\ref{fig:Pantheon_precision} (left column) and the relative reconstruction imprecisions (right column).

We compare our reconstruction to the respective quantities of a $\Lambda$CDM model as parametrised by \cite{bib:Planck} and by \cite{bib:Scolnic} as summarised in Table~\ref{tab:reference_LCDM}. In order to compare $E(a)$ on equal footage, we normalise it to the value at $a_\mathrm{max}$ for the $\Lambda$CDM models as well. 

For $E(a)$, both model-based reconstructions lie well within the confidence bounds of our reconstruction. The same applies to the model-based reconstructions for $D_\mathrm{L}(a)$ and $D(0.5,z_\mathrm{s})$ as parametrised by \cite{bib:Planck}. As expected, the tension in $H_0$ between \cite{bib:Planck} and \cite{bib:Riess_H0} causes the $D_\mathrm{L}(a)$ and $D(0.5,z_\mathrm{s})$ as determined by the parametrisation of \cite{bib:Scolnic} to lie below the 99\% confidence bounds of our reconstruction because \com{we employ} $H_0$ as \com{derived} by \cite{bib:Planck}. 

\subsubsection{Fitting a flat $\Lambda$CDM model as consistency check}

\comm{Not finding any significant tensions with \cite{bib:Planck} and \cite{bib:Scolnic}, we perform an additional consistency check and employ our $\chi^2$-parameter-estimation function defined by Equation~\eqref{eq:chi2} to fit a flat $\Lambda$CDM to the luminosity distance data of the Pantheon sample. We use the average standard absolute magnitude $M=-19.25$ from local supernovae as determined in \cite{bib:Richardson} to scale the $\tilde{D}_{\mathrm{L},i}$ in Equation~\eqref{eq:D_Lred}. Subsequently, we insert the luminosity distance data of the Pantheon and the luminosity distance measure as given by Equation~\eqref{eq:DL_LCDM} and parametrised by $H_0$ and $\Omega_{\mathrm{m}0}$ ($\Omega_\Lambda=1-\Omega_{\mathrm{m}0}$) into Equation~\eqref{eq:chi2}. The resulting non-linear least-squares optimisation problem can be solved by the standard \emph{lsqcurvefit} routine in MATLAB. As $M$ and $H_0$ are arbitrary but dependent scales, we vary the value of $M$ to find that  it only changes the fitted value of $H_0$  and leaves the resulting value for $\Omega_{\mathrm{m}0}$ invariant, as we expected. We obtain $\Omega_{\mathrm{m}0}=0.2870$ at $\chi^2_\nu=0.9084$. To determine confidence bounds on this value, we use the same 1000 Monte-Carlo-simulated Pantheon-like samples (see Section~\ref{sec:confidence_bounds}) that we used to determine the confidence bounds for our approach. Fitting these samples to the $\Lambda$CDM model set up above, we obtain for the median and the subsequent 68\%, 95\%- and 99\% confidence bounds
\begin{equation}
\Omega_{\mathrm{m}0} = 0.287 \pm \phantom{.}^{0.300}_{0.275} \pm  \phantom{.}^{0.312}_{0.262} \pm \phantom{.}^{0.330}_{0.246} \;.
\end{equation} 
Thus, our fit agrees with \cite{bib:Scolnic} within the 68\% confidence bound and with \cite{bib:Planck} within the 95\% confidence bound.}

\subsection{Reconstruction accuracy from simulated data}
\label{sec:accuracy}

Having determined the reconstruction precision for the Pantheon sample, we investigate the reconstruction accuracy for a Pantheon-like simulated data set \comm{in a flat $\Lambda$CDM model cosmology}, as detailed in Section~\ref{sec:data}. All remaining input to the reconstruction is taken from the specifications listed at the end of Section~\ref{sec:basis_selection}. Figure~\ref{fig:Pantheon_accuracy} shows the results. The reconstruction of $\Lambda$CDM (\cite{bib:Planck}) (or of \cite{bib:Scolnic}) is subtracted from our reconstruction.


The plot on the left of Figure~\ref{fig:Pantheon_accuracy} shows the increase in accuracy for $E(a)$ with an increasing amount of basis functions. For $N_\mathrm{B}=3,4$, the closed-form basis functions are used, while for $N_\mathrm{B}=5$, we employ the numerical implementation. \comm{Using the numerical implementation of $N_\mathrm{B}=5$ for $N_\mathrm{B}=3,4$ as well, differences between the analytic and the numerical reconstructions of $E(a)$ only differ on the order of $0.001$, i.e.\@ invisibly in this plot.} While reconstructions with $N_\mathrm{B}=3,4$ favour the parametrisation of \cite{bib:Scolnic}, for $N_\mathrm{B}=5$ the parametrisation of the underlying simulation of \cite{bib:Planck} is finally preferred. 

Reconstructing $D_\mathrm{L}(a)$ with $N_\mathrm{B}=4$, we compare the implementation with the numerical and the analytic basis functions in the central plot of Figure~\ref{fig:Pantheon_accuracy}. \comm{To test the impact of the implementation on the reconstruction accuracy, we employ two different numerical implementations: one which only assumes $N_\mathrm{S}=1048$ sampling points for each basis function at the same scale factors as the Pantheon sample. The second one, also used in Section~\ref{sec:basis_selection}, assumes $N_\mathrm{S}=10000$ equidistantly distributed sampling points of $a \in \left[a_\mathrm{min}, 1\right[$ for each basis function.} 

For small scale factors, the 1\% gain in accuracy between the numerical implementation sampled at the scale factors of the Pantheon sample (i.e.\ $\Phi$ is evaluated at $N_\mathrm{S}=1048$ sampling points) and the analytic basis functions can also be achieved for a numerical basis function with $N_\mathrm{S} = 10000$ sampling points. Yet, this increases the run-time of the $D_\mathrm{L}$-routine by more than a factor of 4 and still causes numerical instabilities when the scale factor approaches 1.

For the lensing distance ratio $D$, determined for a typical lens at redshift $z_\mathrm{l}=0.5$ as a representative example, we obtain the relative inaccuracies as shown in the plot on the right-hand side of Figure~\ref{fig:Pantheon_accuracy}. It shows that the reconstruction with numerical basis functions is slightly worse than the one employing \com{analytic} basis functions. In addition, we plot the relative reconstruction inaccuracies that arise when we first reconstruct $\tilde{D}_\mathrm{L}(a,\hat{\boldsymbol{c}})$ with Equation~\eqref{eq:series_expansion_red}, insert the $M$ as scaling factor that the simulated data has been previously divided by, and subsequently determine $D$ from the $D_\mathrm{L}(a,\hat{\boldsymbol{c}})$ (using Equation~\eqref{eq:D_A} to convert $D_\mathrm{L}$ to $D_\mathrm{A}$ to be inserted into Equation~\eqref{eq:D}). As stated in Section~\ref{sec:observational_properties}, using a global measurement of $M$ as scaling would thus yield a higher accuracy for sources close to the lens (for $z_\mathrm{l}=0.5$ the sources should be located between $z_\mathrm{s}=0.5$ and 1).



\begin{figure*}
\centering
  \includegraphics[width=0.51\textwidth]{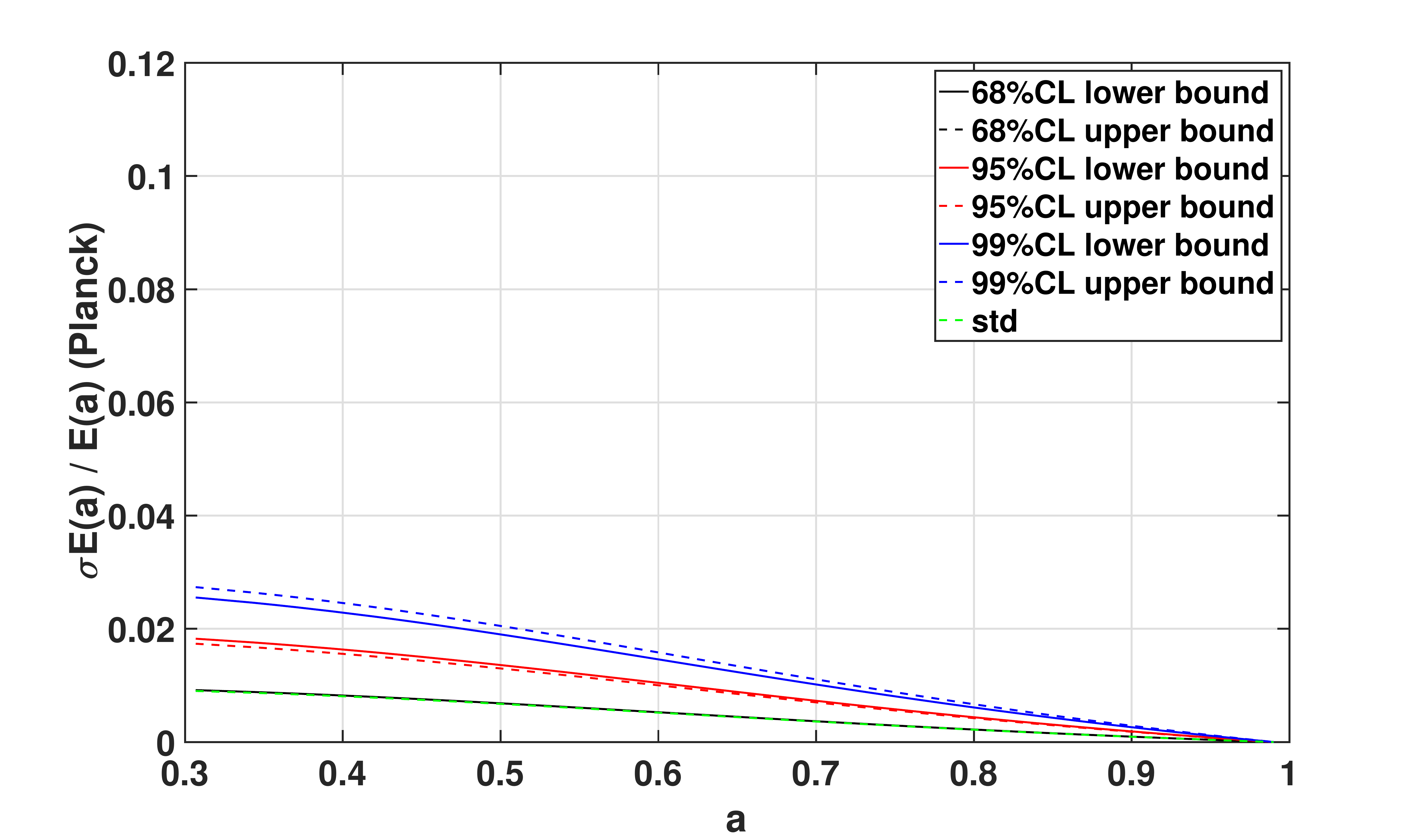} \hspace{-0.04\textwidth}
  \includegraphics[width=0.51\textwidth]{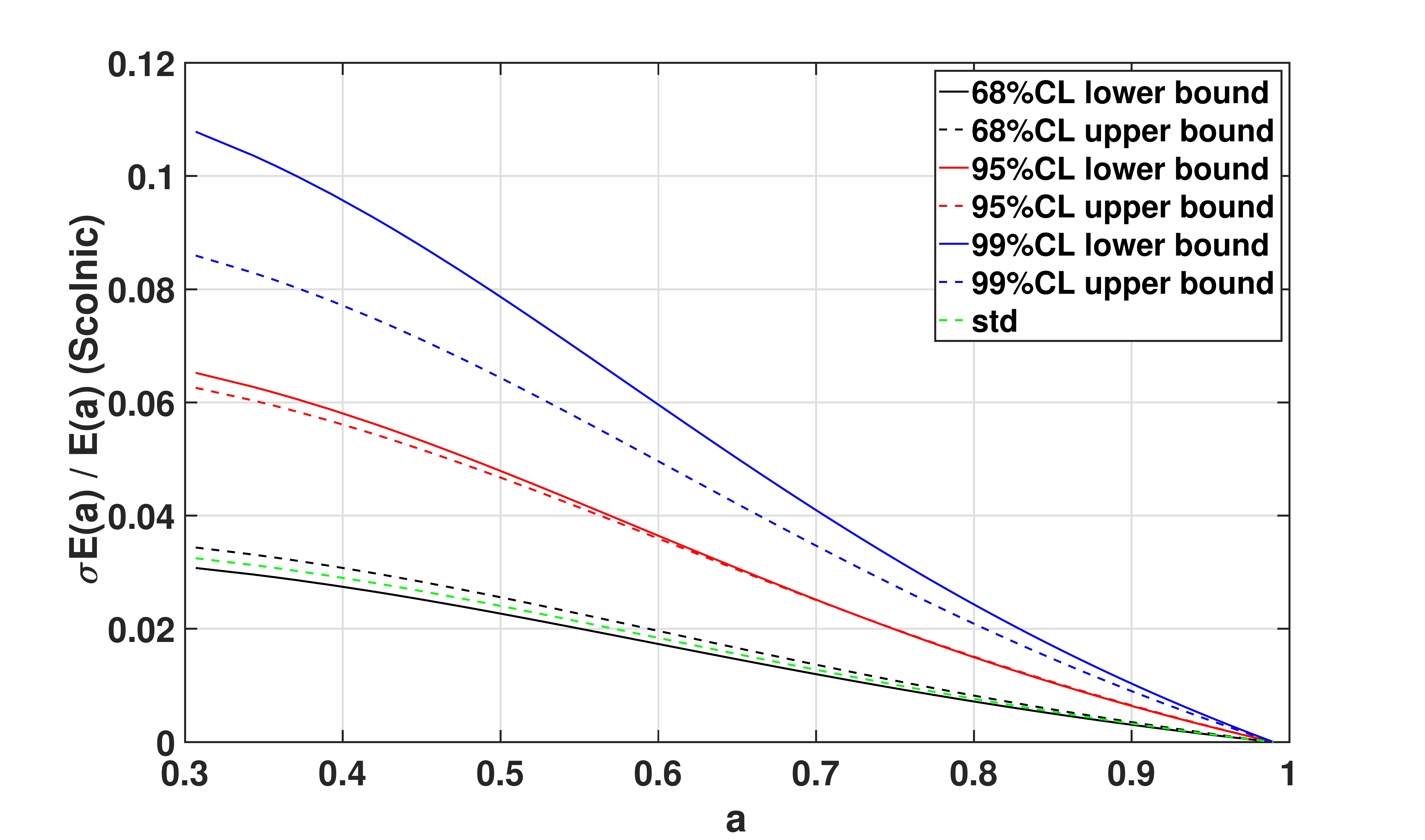} \\
 \includegraphics[width=0.51\textwidth]{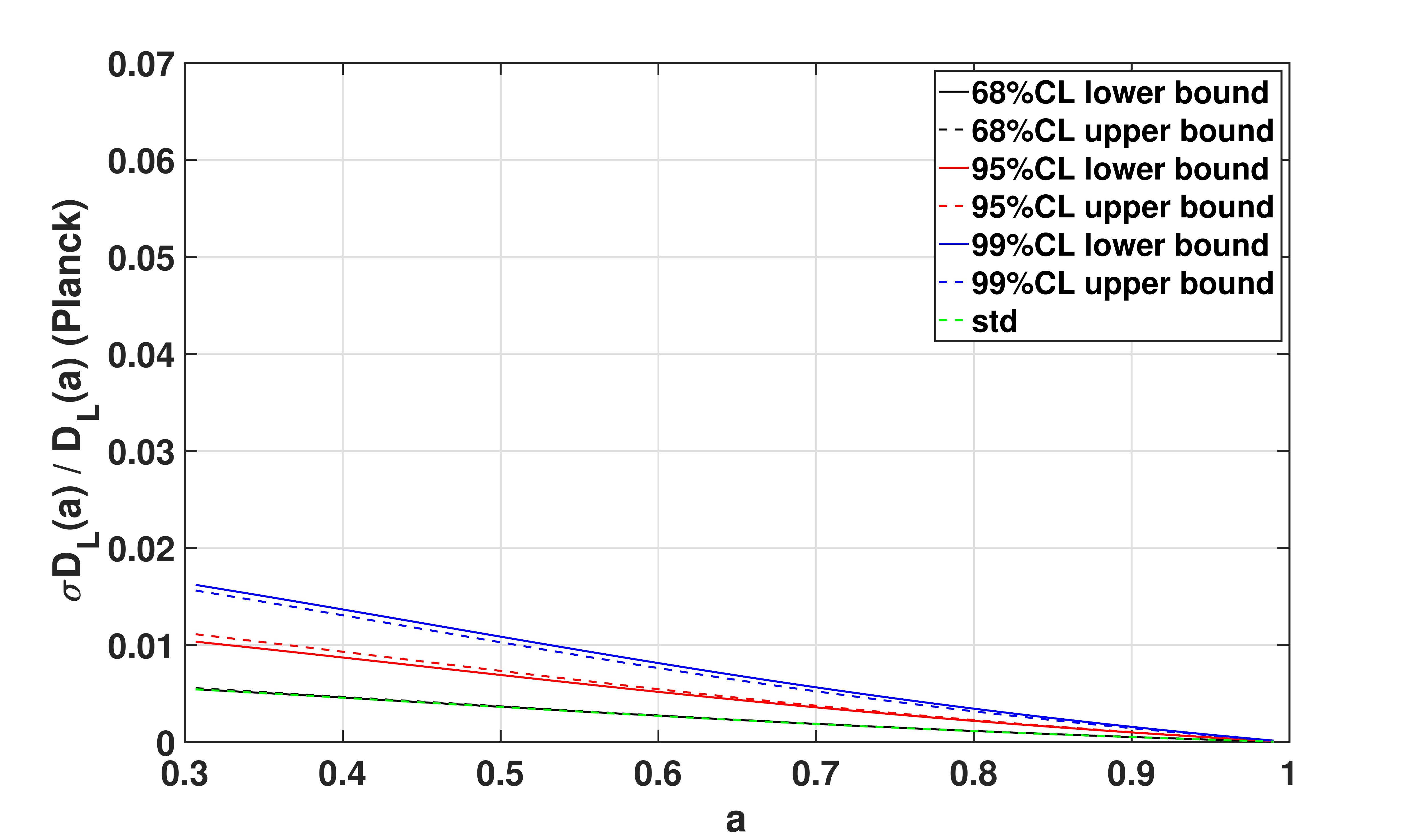} \hspace{-0.04\textwidth}
  \includegraphics[width=0.51\textwidth]{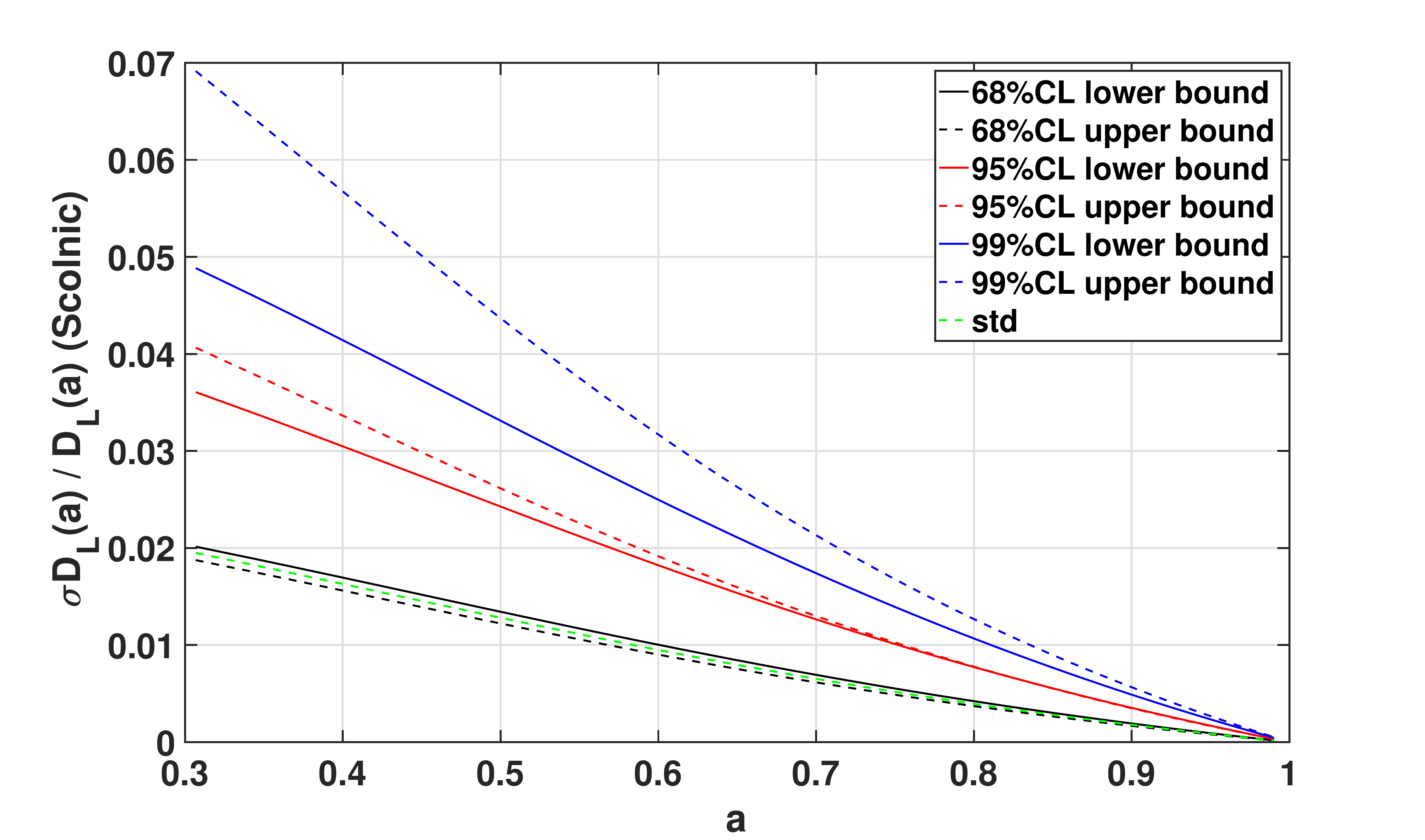} \\
 \includegraphics[width=0.51\textwidth]{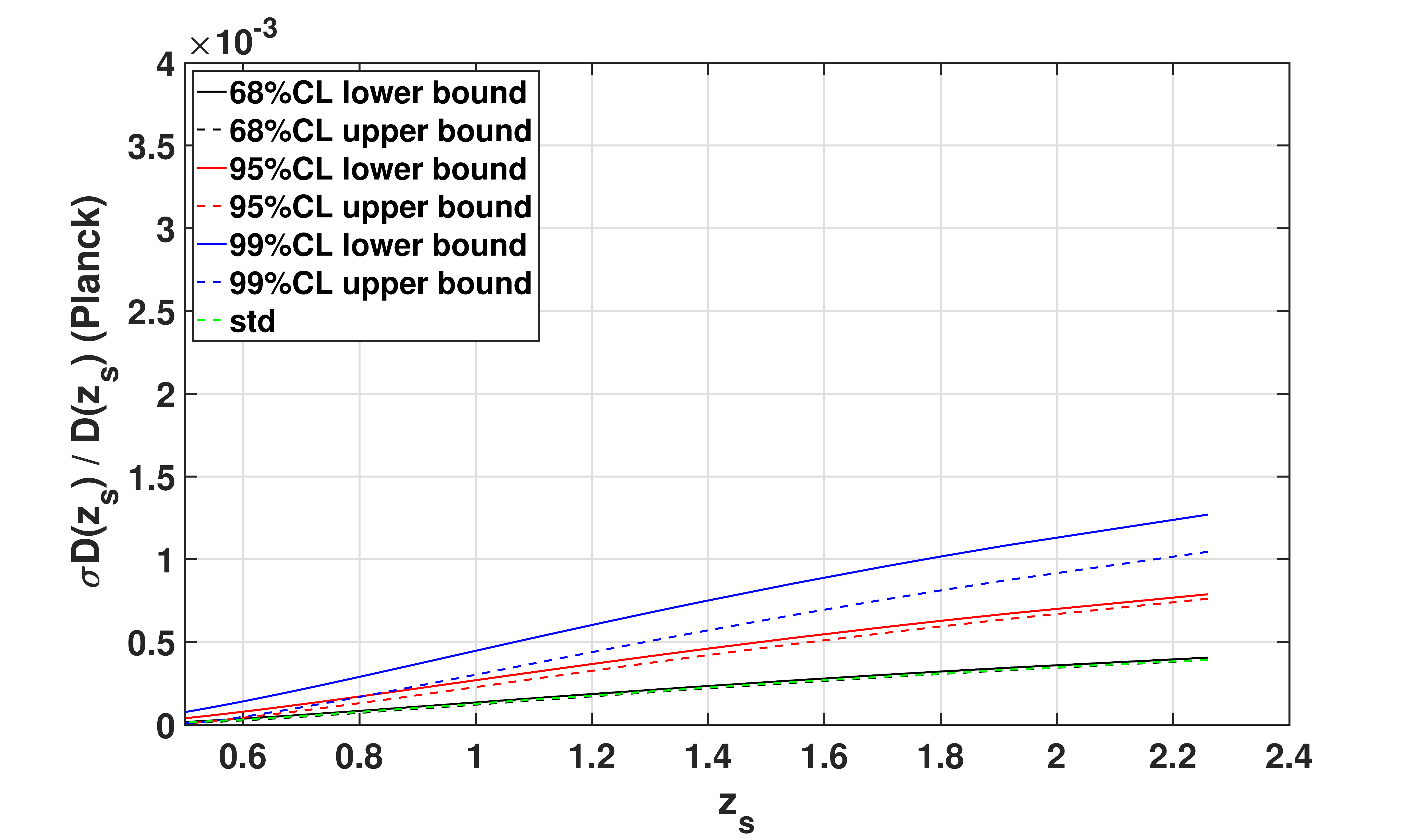} \hspace{-0.04\textwidth}
  \includegraphics[width=0.51\textwidth]{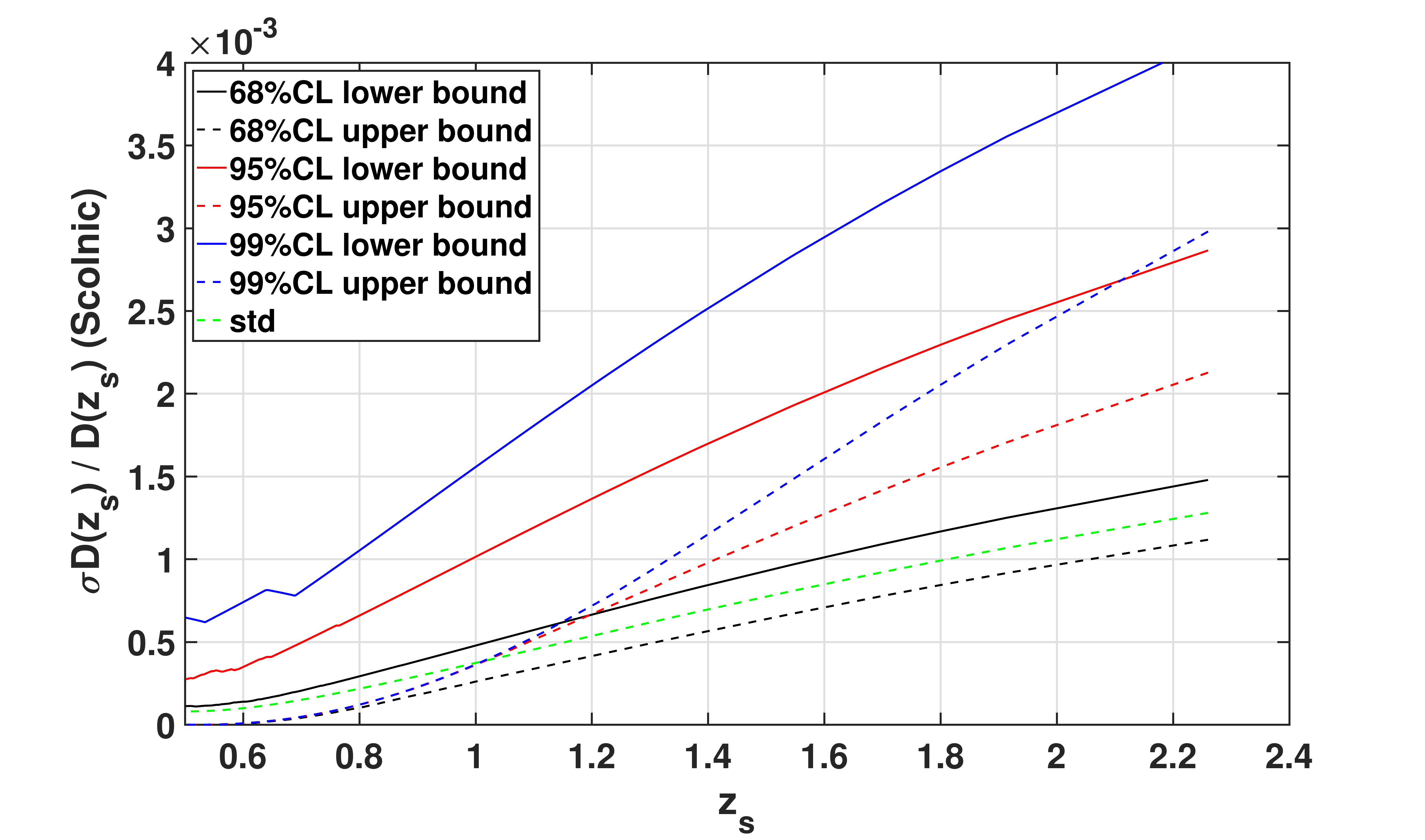}
   \caption{Left column: Relative imprecisions of $E(a)$, $D_\mathrm{L}(a)$, and $D(0.5,z_\mathrm{s})$ for a parametrisation of the Friedmann model according to \protect\cite{bib:Planck}, Right column: Same plots for a parametrisation according to \protect\cite{bib:Scolnic}. \com{The confidence bounds are obtained in a Monte-Carlo simulation, generating one Pantheon-like sample data set for 1000 different $\Lambda$CDM models based on the parametrisations by \protect\cite{bib:Planck} or by \protect\cite{bib:Scolnic}.}}
\label{fig:LCDM_precision}
\end{figure*}

\subsection{Comparison to model-based reconstructions}
\label{sec:model_based_reconstructions}

Next, we compare the confidence intervals of our Friedmann-parameter-free reconstructions with the confidence intervals of the model-based reconstructions parametrised by \cite{bib:Planck}, which is the most precise one, and by \cite{bib:Scolnic}, based on the Pantheon sample. The parametrisations are shown in Table~\ref{tab:reference_LCDM}, the confidence intervals for the parameters are summarised in Table~\ref{tab:confidence_LCDM}. 
\begin{table}
\caption{Confidence intervals of $\Lambda$CDM parametrisations of \protect\cite{bib:Planck} (first row) and \protect\cite{bib:Scolnic} (second row) to determine confidence intervals on model-based $E(a)$, $D_\mathrm{L}(a)$, and $D(z_\mathrm{l},z_\mathrm{s})$.}
\begin{tabular}{cccccc}
\hline 
$\Lambda$CDM & $\Delta \Omega_\mathrm{r0}$ & $\Delta \Omega_\mathrm{m0}$ & $\Delta \Omega_\mathrm{K}$ & $\Delta \Omega_{\Lambda}$ & $\Delta H_0$ \\
model & & & & & [km/s/Mpc] \\
\hline
(Planck) & 0.0 & 0.0062 & 0.0 & 0.0062 & 0.46 \\
(Scolnic) & 0.0 & 0.022 & 0.0 & 0.020 & 1.62$^{(1)}$ \\
\hline
\end{tabular}
$^{(1)}$taken from \citet{bib:Riess_H0}
\label{tab:confidence_LCDM}
\end{table}

Since the absolute scale, e.g. $H_0$ (see Section~\ref{sec:observational_properties}), cannot be determined from the supernova sample and is thus subject to the same confidence intervals for all reconstructions, we only focus on the confidence intervals of the $\Omega_i$. 
Evaluating relative uncertainties, i.e.\ $\sigma E(a)/E(a)$, $\sigma D_\mathrm{L}(a)/D_\mathrm{L}(a)$ and $\sigma D(z_\mathrm{s})/D(z_\mathrm{s})$, $H_0$ cancels out. 
To determine confidence bounds on $E(a)$ for the parametrisations according to \cite{bib:Planck} and \cite{bib:Scolnic}, we employ the definition of $E(a)$ in Equation~\eqref{eq:H} and draw 1000 Pantheon-like data sets at the scale factors of the Pantheon sample from a Monte-Carlo simulation \comm{of a flat $\Lambda$CDM model}. 
Each Pantheon-like data set is generated with a different Friedmann parametrisation $(\Omega_{m0}, \Omega_\Lambda=1-\Omega_{m0})$ drawn from a Gaussian distribution around the values listed in Table~\ref{tab:reference_LCDM} with a standard deviation given by the confidence intervals shown in Table~\ref{tab:confidence_LCDM}. 
Confidence intervals for $E(a)$ parametrised by \cite{bib:Planck} and \cite{bib:Scolnic} are then derived from \com{this Monte-Carlo simulation} in the same way as for the Friedmann-parameter-free reconstruction (see Section~\ref{sec:confidence_bounds}). 
Subsequently, the confidence intervals on the $\Lambda$CDM-parametrised $D_\mathrm{L}(a)$ and $D(0.5,z_\mathrm{s})$ are determined.
 Figure~\ref{fig:LCDM_precision} (left) shows the relative imprecision in the reconstructions based on the parametrisation according to \cite{bib:Planck}, Figure~\ref{fig:LCDM_precision} (right) shows the same plots for the reconstructions based on the parametrisation according to \cite{bib:Scolnic}. 

We find that the imprecisions for both $\Lambda$CDM parametrisations are of the same order with tighter confidence bounds for \cite{bib:Planck} for $E(a)$ and $D_\mathrm{L}(a)$, which is expected from the smaller confidence bounds on $\Omega_{m0}$ (see Table~\ref{tab:confidence_LCDM}). 

As we can observe from a comparison of the plots in Figure~\ref{fig:LCDM_precision} with the right-hand side of Figure~\ref{fig:Pantheon_precision}, the precision of $E(a)$ of both model-based reconstructions is about one order of magnitude higher than the precision of $E(a)$ of the Friedmann-parameter-free reconstruction. For $D_\mathrm{L}(a)$, the precision of the model-based reconstructions is about a factor of three higher \comm{for the parametrisation by \cite{bib:Planck} and on equal footing for the parametrisation by \cite{bib:Scolnic}.} For $D(0.5,z_\mathrm{s})$, the model-based reconstruction is about a factor of \comm{20}-50 more precise than our reconstruction.

\section{Synopsis of results compared to other sources of imprecision in lensing}
\label{sec:synopsis}

Summarising the results from Section~\ref{sec:application}, we find that the lensing distance ratio $D$ for a typical lens redshift $z_\mathrm{l}$ can be reconstructed without specialising a parametrisation for the underlying Friedmann model with a relative imprecision of the order of 1-2\% (68\% confidence level), 2.5-5\% (95\% confidence level), and 3.5-8\% (99\% confidence level). Parametrising the Friedmann model with the parameter values of \cite{bib:Planck} or \cite{bib:Scolnic} (see Table~\ref{tab:reference_LCDM}), the relative imprecisions are below \comm{0.41\%} for all three confidence levels.

In the model-independent approach to characterise gravitational lenses as developed in \cite{bib:Wagner1}, \cite{bib:Wagner2}, and \cite{bib:Wagner3} so far, the lensing distance ratio only enters in the time-delay equation, Equation~\eqref{eq:time_delay}, while $D$ does not enter the equations to determine lens properties (locally constrained reduced shear and ratios of potential derivatives) by positions and shapes of multiple images in the lens plane. \com{Therefore, we only have to consider the impact of data-based distances in Equation~\eqref{eq:time_delay}.}


Having measured a time delay between two multiple images, the difference in the lensing potential between those images can be determined with Equation~\eqref{eq:time_delay}. In \cite{bib:Wagner3}, we showed that time-delay measurements between multiple images fix the enclosed mass density for a given cosmological model, i.e. for a known lensing distance ratio $D$. Hence, inserting $E(a)$ as reconstructed by the supernovae \comm{and a given $H_0$} into Equation~\eqref{eq:time_delay}, we can uniquely determine $\Delta \phi$, given the mathematical prerequisites on $\phi$ detailed in \cite{bib:Wagner3} are fulfilled.

Assuming that the distance and time-delay measurements are performed independently, the relative uncertainty of $\Delta \phi$ is given by
\begin{align}
\dfrac{\delta(\Delta \phi)}{\Delta \phi} = \sqrt{\left( \dfrac{\sigma_z}{1+z_\mathrm{l}} \right)^2 + \left( \dfrac{\sigma_D}{D}\right)^2 + \left(\dfrac{\sigma_\tau}{\tau_{ij}} \right)^2} \;.
\end{align}

We \comm{conservatively} estimate the imprecision of the redshift, \comm{$\sigma_z$}, to 1\%, considering spectroscopic analysis, as e.g. performed in \cite{bib:Scodeggio}. \comm{If the redshift of the supernova is not acquired from the supernova itself but from its host, the relative uncertainty can be of the order of 0.1\%, \cite{bib:Scolnic}.} Assuming that the time delay is determined between multiple images of quasars, the relative uncertainty, \comm{$\sigma_\tau$}, amounts to 1-5\%, \comm{see e.g.\@ \cite{bib:Wagner_FRB} for an overview on galaxy-cluster scale and \cite{bib:Liao} for a systematic analysis on galaxy scale.} Compared to these estimates, the relative imprecision of $D$, \comm{$\sigma_D$}, as obtained by our reconstruction is of the same order of magnitude, while the relative imprecision of $D$ obtained by a model-based reconstruction is one order of magnitude smaller, so that we obtain
\begin{align}
\left(\dfrac{\delta(\Delta \phi)}{\Delta \phi}\right)_{PF} &\approx \sqrt{\left( 0.01 \right)^2 + \left( 0.01 \right)^2 + \left( 0.01 \right)^2} \approx 1.7\% \;, \\
\left(\dfrac{\delta(\Delta \phi)}{\Delta \phi}\right)_{PB} &\approx \sqrt{\left( 0.01 \right)^2 + \left( 0.001 \right)^2 + \left( 0.01 \right)^2} \approx 1.4\% \;,
\end{align}
where PF and PB stand for our Friedmann-parameter-free (PF) reconstruction and the parametrised Friedmann model (PB), respectively. Hence, generalising the reconstruction of the gravitational lensing potential by not specifying a parametrisation for the Friedmann model, the precision only deteriorates by 0.3\%. \comm{This is the loss in precision that is caused by dropping the model assumption of individual parameters $\Omega_i$ in Equation~\eqref{eq:H_sm}.}

Using a lens model, it is often not necessary to measure a time delay to constrain the mass density profile. Fitting the observables to the lens model already fixes the mass density profile in frequently occurring cases or additional observables as the velocity dispersions along the line of sight are employed, such that $\Delta \phi$ is determined from the lens model with its most-likely model parameters. Further details on this issue, including an analysis where $D$ enters in lens reconstructions with lens models, can be found in the follow-up paper, when we discuss different possibilities to determine $H_0$ from Equation~\eqref{eq:time_delay} and their degeneracies. \comm{State-of-the-art estimates for the currently achievable accuracy and precision for $\Delta \phi$ by lens models can be found in the paper series by \cite{bib:Suyu}.}

\section{Conclusion}
\label{sec:conclusion}

We investigated to which precision and accuracy it is possible to determine the lensing distance ratio $D = D_\mathrm{l}D_\mathrm{s} / D_\mathrm{ls}$ in a generalised Friedmann universe being agnostic about its constituents and their individual contributions. Using the latest compilation of supernovae, the Pantheon sample (\cite{bib:Scolnic}), $D$ can be reconstructed for any combination of lens and source redshifts in the range of 0 to 2.3 with a relative uncertainty on the order of percent. To arrive at this result, we expanded the luminosity distance into a set of orthonormal basis functions. Due to the limited amount of supernovae and their uncertainties, relative inaccuracies in the reconstructed $D$ arise, which \comm{most probably} lie within the 68\% confidence bounds of the measurement uncertainties when we use an Einstein-de-Sitter basis, as introduced in \cite{bib:Mignone}, with 4 basis functions. Compared to three other orthonormal basis sets obtained from analytic functions, the Einstein-de-Sitter basis was shown to be the optimal basis set: It has the tightest confidence bounds that encompasses the relative inaccuracies \comm{of a flat $\Lambda$CDM model simulation} and shows only a slight overfitting in the reconstruction.

\comme{As any other method to reconstruct distances based on standardised, precompiled ensembles of supernovae, our approach also relies on the methods used to fit and calibrate the light curves, which might still be dependent on a cosmological background model and which are based on additional assumptions like the one that uncertainties in the redshift estimates can be incorporated in the uncertainty of the distance modulus (see Section~\ref{sec:theoretical_derivations} for further details). Consequently, improvements in the standardisation process imply changes in our reconstruction method, as we do not work on the raw data directly. For instance, taking into account the redshift uncertainties separately requires our optimisation problem to reconstruct the luminosity distances to change from a least-squares formulation to a total-least-squares formulation. The determination of confidence bounds in our approach can be greatly improved, if the publicly available SNe compilations additionally provide the luminosity distances with their uncertainties.}

Compared to model-based reconstructions of $D$ based on the parametrisations of the $\Lambda$CDM model by \cite{bib:Planck} and \cite{bib:Scolnic}, our Friedmann-parameter-free reconstruction is a factor of \comm{20}-50 less precise than the model-based reconstructions for typical lens and source redshifts between redshifts of 0.5 and 1.0. \comm{This loss can be considered as an estimate for the impact of the cosmological model on $D$, as similarly investigated in \cite{bib:Williams} for the influence of lens model assumptions on the determination of $H_0$ from supernova Refsdal.} Propagating the uncertainties through the time-delay equation to determine the difference in the lensing potential between the two multiple images, $\Delta \phi$, we find that $\Delta \phi$ is less than 0.5\% less precise for the Friedmann-parameter-free reconstruction than for a $\Lambda$CDM model by \cite{bib:Planck} or \cite{bib:Scolnic}. This holds for the typical redshift and time-delay uncertainties on the order of percent that are assumed to be the same for the Friedmann-parameter-free and the parametrised reconstructions. 

Thus, for the model-independent characterisation of gravitational lensing configurations, we conclude that dropping the parametrisation of the Friedmann model in the lensing distance ratio $D$ in favour of a data-based reconstruction allows us to greatly generalise the method at the cost of a small and tolerable additional imprecision. As a drawback, configurations with redshifts larger than 2.3 cannot benefit from Friedmann-parameter-free lensing distance ratios yet and may require the combination of several cosmic probes.


The usage of the reconstructed $E(a)$, $D_\mathrm{L}(a)$, $D_\mathrm{A}(a)$ is not limited to applications in strong and weak gravitational lensing, distances that are determined from a data-based expansion function can also be employed in any kind of astrophysical context. Furthermore, the Friedmann-parameter-free reconstruction of $E(a)$ can be used to determine the linear growth factor, as done in \cite{bib:Haude}, and both, $E(a)$ and the linear growth factor are extensively used in the recently developed approach by \cite{bib:Bartelmann} to set up propagators for particles moving in the phase space of an expanding universe.


A parametrised Friedmann model yields a higher reconstruction precision, yet, there is a plethora of possible parametrisations that cannot be excluded by current observations, e.g. different types of dark energy or modified gravity models, as summarised e.g. in \cite{bib:Amendola} and investigated, e.g. in \cite{bib:Benitez12} \comme{and \cite{bib:Moews}}. We have already analysed the analogous situation in the previous papers of this series for the case of specifying a gravitational lens model. Hence, when considering specific parametrisations of Friedmann models, we should -- analogously to specifying gravitational lens models -- marginalise over all possible parametrisations to obtain the confidence bounds that represent our knowledge most realistically. Those confidence bounds should be compared with the ones obtained by the Friedmann-parameter-free method.

\section*{Acknowledgements}

We would like to thank Matthias Bartelmann, Sophia Haude, Bettina Heinlein, Bruno Leibundgut, Henrik Nersisyan, Dan Scolnic, and the Galaxy Cluster Group at the Institute for Theoretical Astrophysics for helpful discussions and comments. JW gratefully acknowledges the support by the Deutsche Forschungsgemeinschaft (DFG) WA3547/1-3. SM gratefully acknowledges the support by the Deutsche Forschungsgemeinschaft (DFG) BA 1359/20-1.




\bibliographystyle{mnras}
\bibliography{mnras} 



\appendix

\section{Validity restrictions for the series expansion}
\label{app:validity}

Searching for a $D_\mathrm{L}(a,\boldsymbol{c})$ that best fits a given set of measured data at known scale factor positions $a_i$, $i=1,...,N_\mathrm{D}$, can be treated as a sampling problem. Since the scale factor positions belong to a limited interval, the Fourier transform of the data is limited in bandwidth. Furthermore, we assume a random, not necessarily equidistant, spacing of the $a_i$. Then, a generalised version of the Nyquist-Shannon sampling theorem, \cite{bib:Landau}, states that the function underlying the $D_{\mathrm{L},i}$ can be exactly reconstructed, if the average sampling rate is (at least) twice the occupied bandwidth of the signal. 

In our case, the bandwith amounts to $2\pi(1/a_\mathrm{min} - 1/a_\mathrm{max}) \approx 15$, which is much smaller than the average sampling rate of $N_\mathrm{D}/(a_\mathrm{max}-a_\mathrm{min}) \approx 1500$. Hence, for infinitely many coefficients $c_\alpha$ in Equation~\eqref{eq:series_expansion}, the luminosity distance can be uniquely determined, while, in practice, the precision of the measured $D_{\mathrm{L},i}$ limits the number of coefficients, $N_\mathrm{B}$, that can be determined.

\section{Derivation of $\Sigma_{ij}$}
\label{app:sigma_derivation}

We determine the uncertainty on $D_{\mathrm{L},i}$ as
\begin{align}
\Delta D_{\mathrm{L},i} \equiv 10^{\tfrac{\mu + \Delta \mu_i }{5}+1} - 10^{\tfrac{\mu}{5}+1} = D_{\mathrm{L},i} \left( 10^{\tfrac{\Delta \mu_i}{5}} - 1\right) \;,
\end{align}
such that 
\begin{align}
\Delta D_{\mathrm{L},i} \Delta D_{\mathrm{L},j} =  D_{\mathrm{L},i} D_{\mathrm{L},j} \left( 10^{\tfrac{\Delta \mu_i}{5}} - 1\right) \left( 10^{\tfrac{\Delta \mu_j}{5}} - 1\right) \\
= D_{\mathrm{L},i} D_{\mathrm{L},j} \left( 10^{\tfrac{\Delta \mu_i + \Delta \mu_j}{5}} - 10^{\tfrac{\Delta \mu_i}{5}} - 10^{\tfrac{\Delta \mu_j}{5}} + 1 \right) \;. \label{eq:sigma_derivation}
\end{align}
In order to identify $\Delta D_{\mathrm{L},i} \Delta D_{\mathrm{L},j}$ with the symmetric matrix $\Sigma_{ij}$, we symmetrise the right-hand side by identifying
\begin{align}
\Delta \mu_i \equiv \dfrac{\Sigma_{\mu,ij}}{\sqrt{\Sigma_{\mu,jj}}} \; \quad \text{and} \quad \Delta \mu_j \equiv \dfrac{\Sigma_{\mu,ij}}{\sqrt{\Sigma_{\mu,ii}}} \;.
\label{eq:mus}
\end{align}
Inserting Equation~\eqref{eq:mus} into Equation~\eqref{eq:sigma_derivation}, Equation~\eqref{eq:sigma_ij} is obtained. 

\section{$E(a)$ for $K \ne 0$}
\label{sec:k}

For a given curvature $K \ne 0$, we perform the following transformations:
\begin{align}
\tilde{K} = \dfrac{K}{|K|} \;, \quad \tilde{a} = \dfrac{a}{\sqrt{|K|}} \;, \quad \tilde{r} = \sqrt{|K|} r \;.
\label{eq:trafo_k}
\end{align}
Starting from Equation~\eqref{eq:D_L}, the analogous equation to Equation~\eqref{eq:integral_equation1} for $K \ne 0$ is set up in the transformed coordinates using
\begin{align}
\sqrt{|K|} \int \limits_a^1 \dfrac{\mathrm{d}x}{x^2 E(x)} &= \sqrt{|K|} \int \limits_{\tilde{a}}^{\sqrt{|K|}^{-1}} \mathrm{d}\tilde{x} \dfrac{\mathrm{d}x}{\mathrm{d}\tilde{x}} \dfrac{1}{|K| \,\tilde{x}^2 E(\tilde{x})} \\
&= \int \limits_{\tilde{a}}^{\sqrt{|K|}^{-1}} \dfrac{\mathrm{d}\tilde{x}}{\tilde{x}^2 E(\tilde{x})} \;,
\end{align}
and 
\begin{align}
f_{\tilde{K}}(\tilde{r}) = \left\{ \begin{matrix} \sinh(\tilde{r}) & \text{for} \; \tilde{K} = -1 \\ \sin(\tilde{r}) & \text{for} \; \tilde{K} = +1 \end{matrix} \right. \;.
\label{eq:f_tildeK}
\end{align} 
It reads
\begin{align}
 \int \limits_{\tilde{a}}^{\sqrt{|K|}^{-1}} \dfrac{\mathrm{d}\tilde{x}}{\tilde{x}^2 E(\tilde{x})} = f^{-1}_{\tilde{K}}\left( \dfrac{H_0}{c} |K| \,\tilde{a} D_\mathrm{L}(\tilde{a}) \right) \equiv f^{-1}_{\tilde{K}}\left(\tilde{a} \dfrac{D_\mathrm{L}(\tilde{a})}{R_0} \right) \;,
\end{align}
in which we introduced the new scaling $R_0 = c/(H_0|K|)$. Deriving this equation with respect to $\tilde{a}$,
\begin{align}
-\dfrac{1}{\tilde{a}^2 E(\tilde{a})} =\dfrac{\partial f^{-1}_{\tilde{K}}\left(\tilde{a} \dfrac{D_\mathrm{L}(\tilde{a})}{R_0} \right)}{\partial \left( \tilde{a} \dfrac{D_\mathrm{L}(\tilde{a})}{R_0} \right)} \left( \dfrac{D_\mathrm{L}(\tilde{a})}{R_0} + \dfrac{\tilde{a}}{R_0} \dfrac{\mathrm{d}D_\mathrm{L}(\tilde{a})}{\mathrm{d}\tilde{a}} \right)
\label{eq:integral_equation3}
\end{align}
and solving for $E(\tilde{a})$, we arrive at
\begin{align}
E(\tilde{a}) = - \left[ \tilde{a}^2 \dfrac{\partial f^{-1}_{\tilde{K}}\left( \tilde{a} \dfrac{D_\mathrm{L}(\tilde{a})}{R_0} \right)}{\partial \left( \tilde{a} \dfrac{D_\mathrm{L}(\tilde{a})}{R_0} \right)} \left( \dfrac{D_\mathrm{L}(\tilde{a})}{R_0} + \dfrac{\tilde{a}}{R_0} \dfrac{\mathrm{d}D_\mathrm{L}(\tilde{a})}{\mathrm{d}\tilde{a}} \right) \right]^{-1} \;.
\label{eq:E_k}
\end{align}
Hence, employing the transformations of Equation~\eqref{eq:trafo_k}, we obtain $E(\tilde{a})$ as determined by Equation~\eqref{eq:E_k}, in which $D_\mathrm{L}(\tilde{a})$ is scaled by $R_0$, so that we can proceed with the overall normalisation as described in Section~\ref{sec:e}. 
The subsequent steps of the method remain the same as for $K = 0$, taking into account the transformation of Equation~\eqref{eq:trafo_k} and $f_{\tilde{K}}$.

\section{Comparison of the Monte-Carlo simulation and the Fisher-matrix method}
\label{app:Cramer_Rao}

Given the expression for $\chi^2$ in Equation~\eqref{eq:chi2_red}, the Fisher information matrix is determined by
\begin{align}
\mathcal{I}(\boldsymbol{c}) = \Phi^\top \tilde{\Sigma}^{-1} \Phi \quad \in \mathbb{R}^{N_\mathrm{B} \times N_\mathrm{B}} \;.
\end{align}
Given the unbiased estimator $\hat{\boldsymbol{c}}$ for the vector of coefficients (see Equation~\eqref{eq:MLE}), the Cramér-Rao bound states that the covariance of $\hat{\boldsymbol{c}}$, $\text{cov}(\hat{\boldsymbol{c}})$, is bounded from below as
\begin{align}
\text{cov}(\hat{\boldsymbol{c}}) \ge \mathcal{I}^{-1}(\boldsymbol{c}) \;,
\end{align}
which means that $(\text{cov}(\hat{\boldsymbol{c}}) - \mathcal{I}^{-1}(\boldsymbol{c}))$ is positive semi-definite. 

Assuming we sample from a multivariate normal distribution of deviations of $\tilde{D}_{\mathrm{L},i}$ from $\tilde{D}_\mathrm{L}(a,\hat{\boldsymbol{c}})$ in our Monte-Carlo simulation, such that
\begin{equation}
\boldsymbol{X} \equiv \boldsymbol{\tilde{D}}_{\mathrm{L}} - \boldsymbol{\tilde{D}}_\mathrm{L}(a,\hat{\boldsymbol{c}}) \sim \mathcal{N}(\boldsymbol{\tilde{D}}_{\mathrm{L}}, \tilde{\Sigma}_{\mathrm{stat}}) \;,
\end{equation}
with the diagonal matrix $\tilde{\Sigma}_{\mathrm{stat}}$. On its diagonal are the variances caused by the measurement uncertainties of the $\tilde{D}_{\mathrm{L},i}$.
The $\text{cov}(\hat{\boldsymbol{c}})$ based on this multivariate Gaussian $\boldsymbol{X}$ is given by
\begin{equation}
\text{cov}(\hat{\boldsymbol{c}}) = B \cdot \tilde{\Sigma}_{\mathrm{stat}} \cdot B^\top
\end{equation}
with
\begin{equation}
B \equiv \left( \phi^\top \tilde{\Sigma}^{-1} \phi \right)^{-1} \left( \phi^\top \tilde{\Sigma}^{-1} \right) \;.
\end{equation}
Reducing the full covariance matrix $\tilde{\Sigma}$ of the data sample to the statistical uncertainties, $\tilde{\Sigma}_{\mathrm{stat}}$, we arrive at 
\begin{align}
\text{cov}(\hat{\boldsymbol{c}}) = \mathcal{I}^{-1}(\boldsymbol{c}) \;.
\end{align}

\bsp	
\label{lastpage}
\end{document}